\documentclass[useAMS,usenatbib]{mn2e}

\usepackage{graphicx}

\newcommand{\swift}{{\it Swift}}
\newcommand{\xmm}{{\it XMM-Newton}}
\newcommand{\rosat}{{\it ROSAT}}
\newcommand{\kepler}{{\it Kepler}}
\newcommand{\fuse}{{\it FUSE}}

\title[\xmm\ observation of MV\,Lyr]{\xmm\ observation of MV\,Lyr and the sandwiched model confirmation}
\author[A. Dobrotka et al.]{
A. Dobrotka$^1$\thanks{E-mail: andrej.dobrotka@stuba.sk},
J.-U. Ness$^2$,
S. Mineshige$^3$
and A.A. Nucita$^{4,5}$\\
$^1$Advanced Technologies Research Institute, Faculty of Materials Science and Technology in Trnava, Slovak University of Technology\\
in Bratislava, Bottova 25, 917 24 Trnava, Slovakia\\
$^2$XMM-Newton Science Operations Center, European Space Astronomy Center, Camino Bajo del Castillo s/n, Urb. Villafranca del\\
Castillo, 28692 Villanueva de la Ca\~nada, Madrid, Spain\\
$^3$Department of Astronomy, Graduate School of Science, Kyoto University, Sakyo-ku, Kyoto 606-8502, Japan\\
$^4$Department of Mathematics and Physics E. De Giorgi, University of Salento, via per Arnesano, CP 193, 73100, Lecce, Italy\\
$^5$INFN, Sez. di Lecce, via per Arnesano, CP 193, 73100, Lecce, Italy\\
}

\begin{document}

\date{Accepted ???. Received ???; in original form \today}

\pagerange{\pageref{firstpage}--\pageref{lastpage}} \pubyear{2017}

\maketitle

\label{firstpage}

\begin{abstract}
We present spectral and timing analyses of simultaneous X-ray and UV observations of the VY\,Scl system MV\,Lyr taken by \xmm, containing the longest continuous X-ray+UV light curve and highest signal-to-noise X-ray (EPIC) spectrum to date. The RGS spectrum displays emission lines plus continuum, confirming model approaches to be based on thermal plasma models. We test the sandwiched model based on fast variability that predicts a geometrically thick corona that surrounds an inner geometrically thin disc. The EPIC spectra are consistent with either a cooling flow model or a 2-T collisional plasma plus Fe emission lines in which the hotter component may be partially absorbed which would then originate in a central corona or a partially obscured boundary layer, respectively. The cooling flow model yields a lower mass accretion rate than expected during the bright state, suggesting an evaporated plasma with a low density, thus consistent with a corona. Timing analysis confirms the presence of a dominant break frequency around log($f$/Hz) = -3 in the X-ray Power Density Spectrum (PDS) as in the optical PDS. The complex soft/hard X-ray light curve behaviour is consistent with a region close to the white dwarf where the hot component is generated. The soft component can be connected to an extended region. We find another break frequency around log($f$/Hz) = -3.4 that is also detected by \kepler. We compared flares at different wavelengths and found that the peaks are simultaneous but the rise to maximum is delayed in X-rays with respect to UV.
\end{abstract}


\section{Introduction}
\label{introduction}

A great variety of objects such as cataclysmic variables (CVs), symbiotic systems, X-ray binaries or active galactic nuclei are powered by a common physical process: accretion. In binaries the process is based on mass loss from a companion star. The transported gas falls towards the central compact object, and in the absence of a strong magnetic field, an accretion disc forms. The central accretor can either be a white dwarf in the case of CVs and symbiotic systems or a neutron star or a black hole in the case of X-ray binaries (see e.g. \citealt{warner1995} or \citealt{frank1992} for a review).

The family of CVs is divided into several subclasses based on characteristic variability patterns. The most common are the dwarf novae showing quasiregular outbursts with durations of several days and appearing on a time scale of 10 - 100 days (see \citealt{warner1995} for review). VY\,Scl systems spend most of their life time in a high state while they sporadically transition into a transient low state. This alternating behaviour can be explained by corresponding changes in accretion rate. The high state in VY\,Scl systems is stable for a relatively long time, while in a dwarf nova it is only temporary (outbursts). The shorter durations of high states in dwarf novae can be explained by the mass accretion rate being unstable in the framework of the disc instability model (see \citealt{lasota2001} for review). Meanwhile, in VY\,Scl systems, the mass accretion rate remains above the critical limit required for stability, explaining the longer duration of high states. The low state events are generated by a sudden drop of mass transfer from the secondary or even a total stop of mass transfer (\citealt{king1998}, \citealt{hessman2000}).

Observations of accreting systems of various kinds suggest that the basic characteristic of accretion is fast stochastic variability (a.k.a. flickering), see, e.g. \cite{mchardy1988}, \cite{bruch2015}, \cite{vaughan2003a}. The remarkable observational similarities suggest that the same physical mechanism is responsible for flickering in all accretion systems (e.g. \citealt{uttley2002}, \citealt{markowitz2003}, \citealt{vaughan2003b}, \citealt{mchardy2004}). Flickering has three basic observational characteristics; 1) linear correlation between variability amplitude and log-normally distributed flux (so called rms-flux relation) observed in a variety of accreting systems such as X-ray binaries or active galactic nuclei (\citealt{uttley2005}), CVs (\citealt{scaringi2012a}, \citealt{vandesande2015}) and symbiotic systems (\citealt{zamanov2015}), 2) time lags where flares reach their maxima slightly earlier in the blue than in the red (\citealt{bruch2015}) and 3) red noise in power density spectra (PDS). The shape of such PDS can be a simple power law (see e.g. \citealt{shahbaz2010}, \citealt{mushotzky2011}), a broken power law (two power law components) with a single break frequency in between (see e.g. \citealt{kato2002}, \citealt{baptista2008}) or a multicomponent PDS with several characteristic break frequencies (see e.g. \citealt{sunyaev2000}, \citealt{scaringi2012b}, \citealt{dobrotka2014}, \citealt{dobrotka2015b}).

For sufficiently detailed studies of the PDS, a high cadence and long, continuous light curves are needed. An ideal opportunity is offered by the \kepler\ mission which allowed us to discover a multicomponent PDS in the optical waveband of the two CVs V1504\,Cyg (\citealt{dobrotka2015b}) and MV\,Lyr (\citealt{scaringi2012b}). The former case shows two or three characteristic break frequencies, while for the latter, four components were found. It is believed that the accretion process from the companion star toward the white dwarf surface is structured and every single structure has its own flow characteristics behaving differently. Therefore, every single characteristic break frequency in a PDS can be a footprint of a separate accretion structure.

Different simulation techniques/models intended for identifying the sources of flickering (related to accretion structure) have been developed so far. A cellular-automaton model\footnote{A cellular automaton model consists of a regular grid of cells. For each cell, a set of adjacent cells is defined relative to the specified cell. An initial configuration is defined by assigning a state for each cell. Step by step, a new configuration is calculated, according to some rule that determines the new state of each cell in terms of the current state of the cell and the states of adjacent cells.} was proposed by \citet{yonehara1997}, where light fluctuations are produced by occasional flare-like events and a subsequent avalanche flow in the accretion disc atmosphere (see \citealt{mineshige1994} for the original idea). Another cellular-automaton model was developed by \citet{pavlidou2001}. It is based on changing the collection of magnetic flux tubes anchored in the disc, transporting angular momentum and driving accretion inhomogeneities. \citet{dobrotka2010} developed a statistical model to simulate flickering based on the simple idea of angular momentum transport between two adjacent concentric rings in the accretion disc via discrete turbulent bodies. The method is based on a geometrically thin disc with a ratio of $H/r < 0.01$ ($H$ is the disc scale height and $r$ is the distance from the centre). Another way of the PDS reproduction was proposed by \citet{ingram2013}. They derived an analytical expression for the fluctuating accretion rate in the disc, where fluctuations are generated at each radius on a local viscous time scale, and overall variability of the accretion rate in the innermost disc region is the product of all the fluctuations produced at all radii. Practically all mentioned models approve the basic idea of variations in the accretion rate that are produced at different disc radii (\citealt{lyubarskii1997}, \citealt{kotov2001}, \citealt{arevalo2006}), i.e. the mass accretion variability at the outer radii is propagating inside and is influencing the variability characteristics of the inner regions. Such process explains the mentioned linear rms-flux relation and log-normal flux distribution.

\citet{scaringi2014} applied the \citet{ingram2013} method to fit the highest PDS break frequency ${\rm log}(f) = -3.01 \pm 0.06$\,Hz detected in the optical PDS of the nova-like system MV\,Lyr (\citealt{scaringi2012b}). The discs in CVs are believed to be geometrically thin and optically thick. During quiescence the mass accretion rate is low, resulting in a truncated disc (see \citealt{lasota2001} for review), i.e. a hole around the white dwarf is formed because of inefficient cooling\footnote{Cooling in an optically thin hot plasma in this case is based mainly on free-free transitions, therefore dependent on the square of the particle number density. The plasma energy is transported toward the disc by electron conduction, and a low concentration of electrons reduces cooling. The underlying disc matter thus heats and evaporates, which increases the local particle concentration until an equilibrium is reached.} generating evaporation of the matter (\citealt{meyer1994}). Instead of the geometrically thin disc, a central geometrically thick ($H/r$ no longer small compared to 1) optically thin corona forms due to evaporation. \citet{scaringi2014} determine a ratio $H/r > 0.1$ with a radius of $0.81^{+0.20}_{-0.14} \times 10^{10}$\,cm and an $\alpha$ parameter of 0.6 - 1.0 (\citealt{shakura1973}). The author interpreted such a disc as an expanded optically thin hot corona surrounding a geometrically thin standard disc, i.e. the so-called sandwiched model.

\citet{dobrotka2015a} applied a modelling method (\citealt{dobrotka2010}) to the study of MV\,Lyr. They find the geometrically thin disc to be responsible for the second lowest break frequency, while the lowest one can be generated by somehow enhanced activity of the outer disc rim. From their modelling, they also find a solution where part of the central disc can be responsible for the highest break frequency fitted by \citet{scaringi2014}. Although the derived disc radius of $10^{10}$\,cm and $\alpha = 0.8 - 1.0$ are in agreement with \citet{scaringi2014}, the ratio $H/r$ is completely different, i.e. $H/r < 0.1$ in \citet{dobrotka2015a} and $H/r > 0.1$ in \citet{scaringi2014}.

The explanation of the optical multicomponent PDS of the nova-like MV\,Lyr promises the most complex image of the accretion process in a CV so far. However, the character of the highest break frequency is still not perfectly clear because of two contradictory geometry conditions from numerical modelling, but the corona interpretation is very plausible. There is only one way to resolve this puzzle, i.e. a direct X-ray observation, because the hot optically thin corona is radiating in X-rays. If the corona interpretation is correct, the highest optical break frequency ${\rm log}(f) = -3.01 \pm 0.06$\,Hz must be detected in the X-ray PDS as well.

While MV\,Lyr has been observed in X-rays before, none of them were sufficient for the required timing analysis. During the \rosat\ all-sky survey, MV\,Lyr was observed three times for 722\,s (in Oct. 1990), 20250\,s (in Nov. 1992), and 2218\,s (in May 1996) with the position-sensitive proportional counter (PSPC) being the first two observations coincident with high optical state periods and the last occurring during a low state. The PSPC $0.1-2.4$\,keV count rates were $0.079 \pm 0.011$, $0.069 \pm 0.002$ and $< 0.0008$ (\citealt{greiner1998}) corresponding (after re-scaling the assumed source distance from 320\,pc to 500\,pc) to luminosities of $4.87 \times 10^{31}$\,erg\,s$^{-1}$, $3.87 \times 10^{31}$\,erg\,s$^{-1}$, and $< 1.20 \times 10^{30}$\,erg\,s$^{-1}$, respectively. In contrast to our \xmm\ observation, the PSPC bandpass does not cover the hard range above 2.4keV that contains a lot of critical information.

MV\,Lyr was also observed by \swift\ during both high (Obs ID 91443 lasting for 14240\,s) and low state (Obs ID 32042, 3282\,s). \citet{balman2014} concentrated on the former and found that the spectrum can be well fitted by a multi-temperature plasma emission model ({\tt CEVMKL} in {\tt xspec}). In order to test the hypothesis of the presence of any scattering effect of X-rays from a wind or an extended component, these authors realized that adding a power-law component increased the quality of the fit. In particular, they derived only a lower limit to the plasma temperature ($kT > 21$,keV), $\Gamma \simeq 0.82$, $\alpha \simeq 1.6$ and hydrogen column density of $0.13 \times 10^{22}$\,cm$^{-2}$. The 0.2-10\,keV flux of $5.4 \times 10^{-12}$\,erg\,s$^{-1}$\,cm$^{-2}$ corresponds to a luminosity of $1.7 \times 10^{32}$\,erg\,s$^{-1}$.

Using \rosat\ data, \citet{balman2014} derived a $2 \sigma$ upper limit to the soft black-body component (if any) associated to the boundary layer finding $kT < 6.6$\,eV. Taking the accretion rates derived by optical and UV bands, the standard disc models predict an optically thick boundary layer with temperature (13 - 33\,eV) larger than that derived from the \rosat\ data and luminosity ($> 10^{34}$\,erg\,s$^{-1}$) much higher than that inferred by using the \swift\ data. Moreover, the X-ray luminosity in the $0.1 - 50$\,keV band of the source is $\simeq 3.2 \times 10^{32}$\,erg\,s$^{-1}$, so that its ratio to the disc luminosity (calculated from the UV and optical observations) is in the range $0.01 - 0.001$ thus suggesting that the MV\,Lyr system in high state is characterized by an optically thin boundary layer associated with a low efficiency accretion model (an ADAF-like flow) and/or an X-ray corona on the inner disc and close to the WD.

\citet{zemko2014} studied all \swift\ XRT spectra, thus the observation taken during low state and the same data studied by \citet{balman2014}. They found that a good fit is obtained (see their Table~5) with a two-component thermal plasma model but also with a thermal plasma to which a power-law is added. In the high state, MV\,Lyr is characterized by a 0.3 - 10\,keV luminosity of $1.7 \times 10^{32}$\,erg\,s$^{-1}$ (consistently with \citealt{balman2014}) which reduces to $3.6 \times 10^{31}$\,erg\,s$^{-1}$ in the low state.

In this paper we present timing and spectral analyses of our recent \xmm\ observation of MV\,Lyr. The data are described in Section~\ref{obs}, the EPIC spectrum is analysed in Section~\ref{spectrum}, and timing analysis in Section~\ref{timing}. We discuss and summarize our results in Sections~\ref{discussion} and \ref{summary}, respectively.

\section{Observations}
\label{obs}

\begin{figure}
\includegraphics[width=75mm,angle=-90]{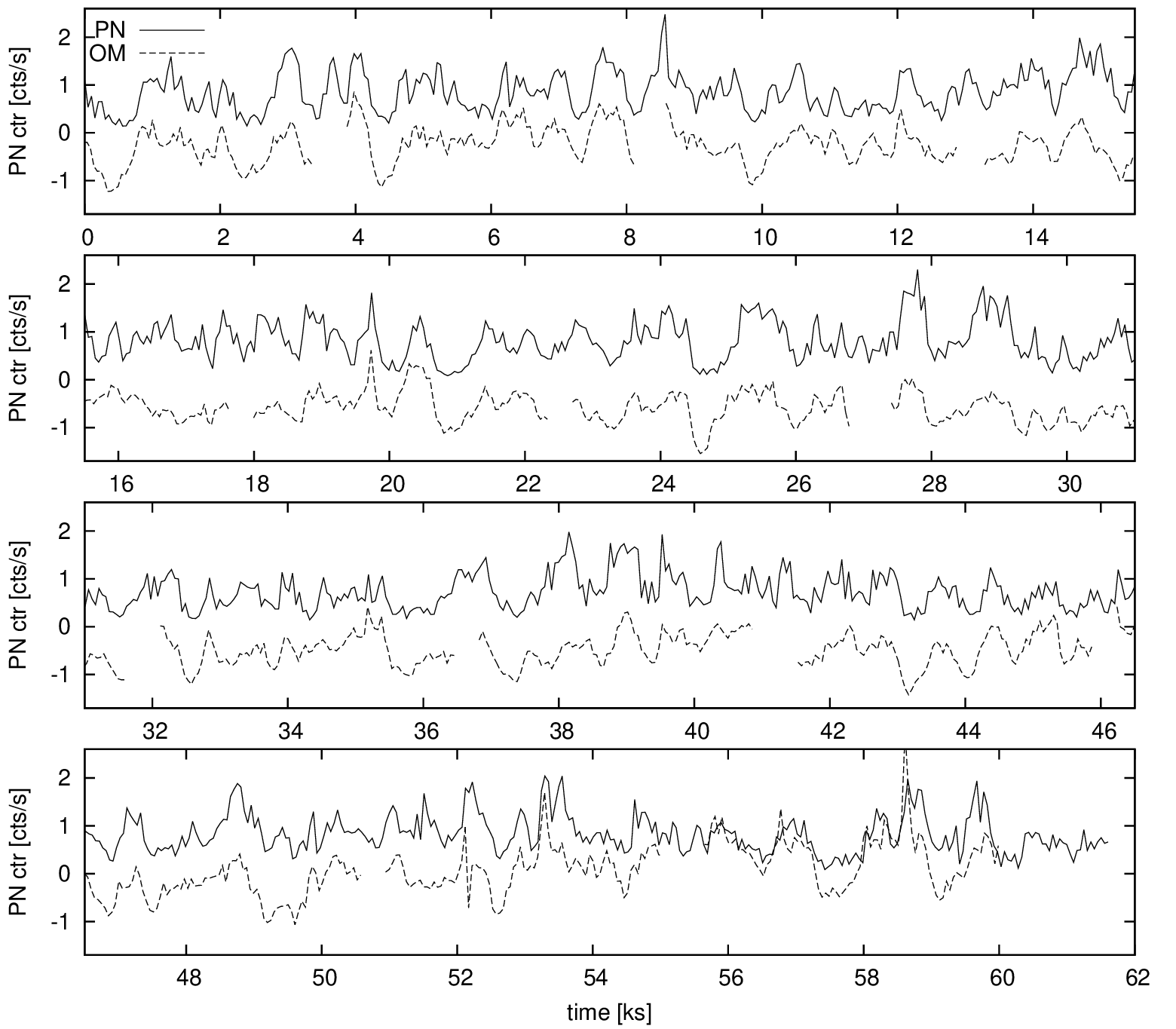}
%\resizebox{\hsize}{!}{\includegraphics[angle=-90]{lc_PN_OM.eps}}
\caption{\xmm\ X-ray (PN, solid line) and UV (OM, dashed line) light curves binned into 50\,s bins for visualization purpose. The OM data are modified (first divided by 100 and subsequently offset by -3.3) in order to make both light curves comparable. Intensity units are count rate (ctr).}
\label{lc}
\end{figure}

\begin{figure}
\resizebox{\hsize}{!}{\includegraphics[angle=-90]{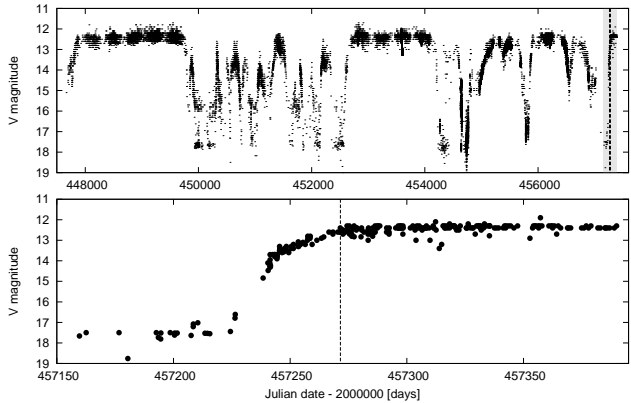}}
\caption{Long-term AAVSO visual light curve of MV\,Lyr. The time when the \xmm\ observation was taken is marked by the vertical dashed line. The lower panel is a detail of the upper panel (shaded area at the end).}
\label{lc_aavso}
\end{figure}

Since the existing \rosat\ and \swift\ X-ray observations either only cover a limited energy range or are too short for the sensitivity of timing analyses needed to put the \kepler\ data into context, we have requested a 50-ks \xmm\ observation. On 2015 September 6, this observation was realised during the more common high state. A 64.6-ks observation was taken under ObsID 0761040101 yielding 61.6, 63.3 and 63.5\,ks for PN, MOS and RGS instruments, respectively. The Optical Monitor (OM) took 13 exposures of 2883s duration each in Image+Fast mode and the UVW1 filter inserted. The data were downloaded from the XMM-Newton Science Archive (XSA), and we obtained science products with the Science Analysis Software (SAS), version 14.0. We used the tool {\tt xmmextractor} to re-generate calibrated events files from which in turn light curves were extracted for all instruments. Optimised extraction regions for the EPIC detectors were calculated with the tool {\tt eregionanalyse} while for the Reflection Grating Spectrometer (RGS) and the optical monitor (OM), standard extraction regions were used by {\tt xmmextractor}. We use the MOS light curves as a comparison while for the timing analysis we use only PN and OM light curves. The PN and OM light curves are shown in Fig.~\ref{lc} and the Julian date of the \xmm\ observation in the long-term context of the AAVSO light curve in Fig.~\ref{lc_aavso}. For the spectral analysis we use both MOS and PN spectra while RGS spectra are only used for consistency checks.

\section{Spectral analysis}
\label{spectrum}

\begin{figure}
\resizebox{\hsize}{!}{\includegraphics{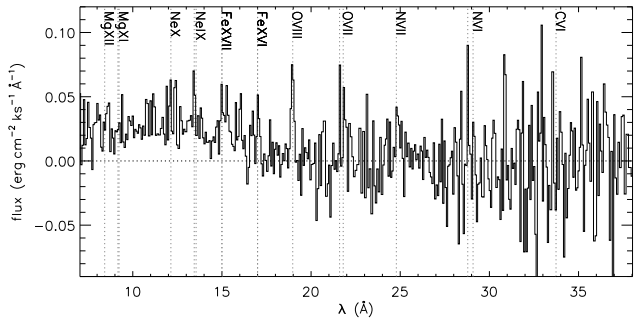}}
\caption{\label{rgsspec}Combined RGS1+2 spectrum in flux units with labels of prominent line transitions.}
\end{figure}

We have extracted X-ray spectra from the RGS and the EPIC cameras using the Science Analysis System (SAS) version 14.0. An RGS spectrum in flux units was obtained using {\tt rgsproc} resulting in a merged spectrum from RGS1 and RGS2, making best use of all redundancies, especially filling chip gaps and bad pixels. The combined RGS spectrum is shown in Fig.~\ref{rgsspec}. Prominent line transitions are marked with vertical lines and labels of the respective ions on top at the respective wavelengths which can be used as a check to see whether any of them are present. Clearly, the RGS spectrum is not sufficiently well exposed for a quantitative analysis, but it allows the conclusion that the spectrum consists of weak continuum below $\sim 17$\,\AA\ (above 0.7\,keV) with superimposed emission lines from oxygen, iron, and neon. Other lines, e.g., N\,{\sc vi} may be present but can not be detected above the noise. The lines are too weak to derive any meaningful line ratios, and, e.g., density diagnostics with He-like triplets can not be applied. But we use the RGS as a consistency check as the high spectral resolution can potentially rule out spectral models, e.g., if they do not reproduce the emission lines that only the RGS can resolve.

For quantitative analysis, we therefore need to rely on low-resolution spectra from the three EPIC detectors PN and MOS1/2 over the energy range 0.1-10\,keV. We merged MOS1 and MOS2 spectra with the SAS tool {\tt epicspeccombine} while we use the PN spectrum separately in a simultaneous fit.
\begin{figure}
\resizebox{\hsize}{!}{\includegraphics{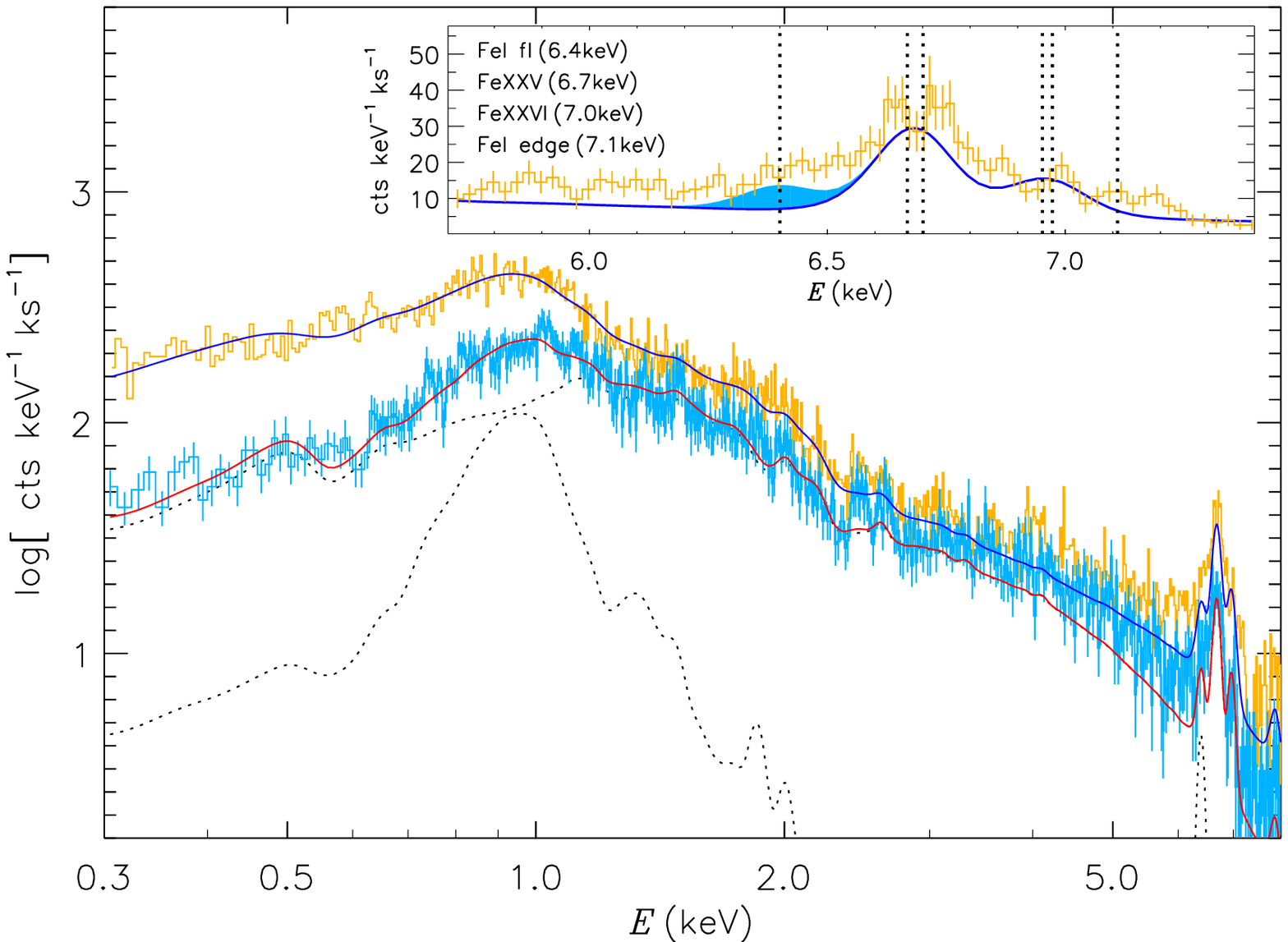}}
\resizebox{\hsize}{!}{\includegraphics{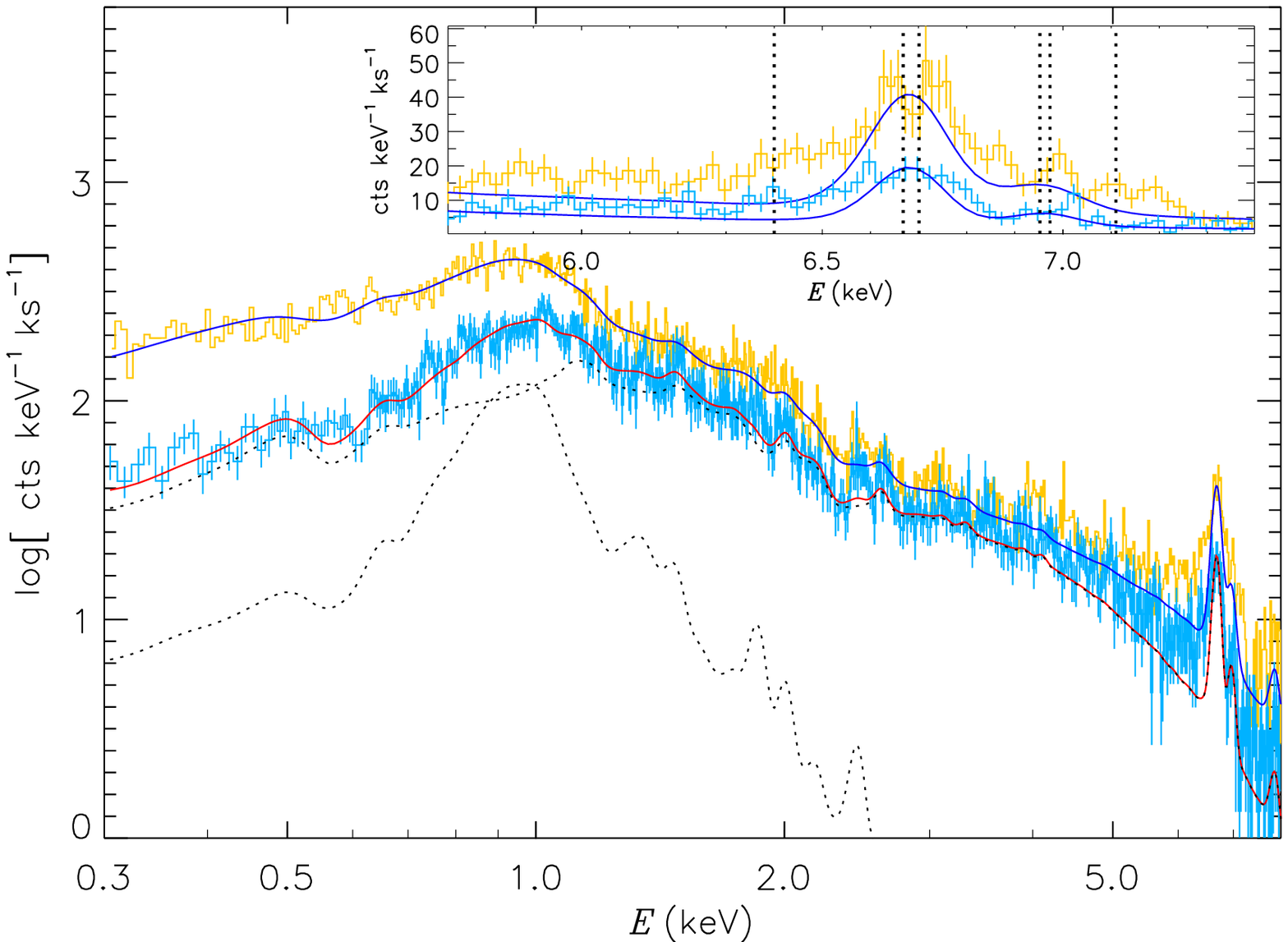}}
\caption{\label{spec}Comparison of models with spectra observed with MOS1+2 (light blue with model in red) and PN (orange with model in dark blue). The model parameters are listed in Table~\ref{models}. The insets show the spectral range between the Fe\,{\sc i} fluorescent line at 6.4\,keV and Fe\,{\sc i} ionisation edge at 7.11\,keV. In addition, the Fe\,{\sc xxv} (6.7\,keV) triplet and Fe\,{\sc xxvi} 1s-2p (6.9\,keV) doublet are marked with vertical dotted lines. {\bf Top}: 2-$T$ {\tt VAPEC} model plus a Gaussian for the Fe\,{\sc i} line at 6.4\,keV (parameters in top part of Table 1). The Fe\,{\sc xxv} and Fe\,{\sc xxvi} lines require an increased Fe abundance relative to solar for the hotter component. The shaded area indicates the resulting Fe\,{\sc i} flux. {\bf Bottom}: Same as above but with a partial absorber applied only to the hot component (middle part of Table~1). Here the Fe lines are well-represented without increasing the Fe abundance.}
\end{figure}
\begin{figure}
\resizebox{\hsize}{!}{\includegraphics{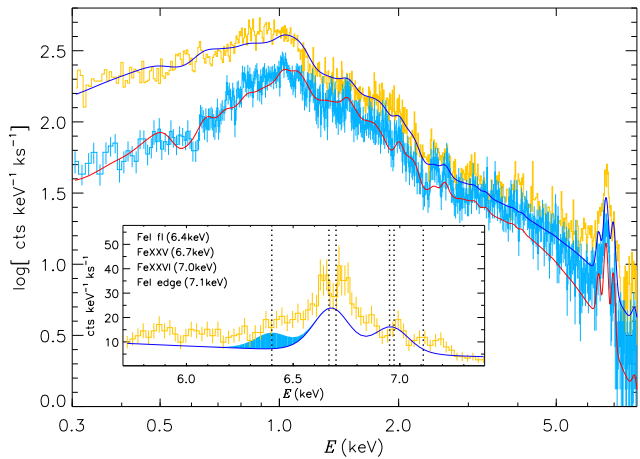}}
\caption{\label{speccflow}Same as Fig.~\ref{spec} with cooling flow ({\tt vmcflow}) model. The  Fe\,{\sc xxv} lines are underpredicted although the Fe abundance has been iterated. Likely, the Fe L complex (around 1\,keV) drives the Fe abundance in the cooling flow model while in the model in the top panel of Fig.~\ref{spec}, with two independent {\tt VAPEC} models, the Fe abundance of the high-T component only affects the Fe\,{\sc xxv} lines.}
\end{figure}

We use {\tt xspec} to test spectral models against the combined MOS1+2 and PN spectra via simultaneous $\chi^2$ minimization fitting. We take photoelectric absorption within the neutral interstellar medium plus potentially circumstellar material into account by the {\tt tbabs} model (T\"ubingen absorption \citealt{wilms2000}). The only parameter of {\tt tbabs} is the neutral hydrogen column density $N_{\rm H}$. As an independent estimate for the amount of interstellar absorption in the direction of MV\,Lyr we use the HEASARC NH tool\footnote{http://heasarc.gsfc.nasa.gov/cgi-bin/Tools/w3nh/w3nh.pl}. Two values are computed from Galactic H\,{\sc i} measurements in the Leiden/Argentine/Bonn (LAB) Survey and by \cite{dickey1990} resulting in values of $5.35\times 10^{20}$\,cm$^{-2}$ and $6.17\times 10^{20}$\,cm$^{-2}$, respectively. During the fit to the X-ray spectra, we keep the parameter $N_{\rm H}$ variable, starting with the interstellar value of $6 \times 10^{20}$\,cm$^{-2}$, in the middle between the above values. If there are any measurable additional sources of circumstellar absorption, a higher value will result, while a lower value may be an indicator for using the wrong model.

In pursuit of finding evidence for either coronal or boundary layer emission, we test three spectral models. All three models are different realizations of the {\tt APEC} model (\citealt{smith2001}), which basically assumes a collisional plasma in equilibrium, thus collisional ionizations and excitations being balanced by radiative recombination and de-excitations. This way a spectrum consisting of bremsstrahlung continuum (from free electrons) plus emission lines (from bound-bound transitions) is produced. The key parameter is the electron temperature $T$ which is converted to a Maxwellian velocity distribution of electrons and ions. A low-density plasma is assumed with an arbitrary density of log($n_e)=1$ (in cgs units), and that no photons are re-absorbed (thus optically thin). The intensity depends on the differential emission measure, basically the product of volume and density, $V\times n_e$. For details we refer to \cite{smith2001}. An isothermal plasma is unlikely in nature, more likely is a broad distribution of temperatures. This can be approximated by a multi-temperature model or by integration over a continuous temperature distribution. A comparison of these two approaches is presented by \cite{ness2009}.
The {\tt APEC} model assumes solar abundances by \cite{agrev89} that can be scaled by a factor to determine an overall metallicity. Meanwhile the {\tt VAPEC} model allows modification of individual abundances. We used the {\tt VAPEC model} iterating the Fe abundance if needed.\\

We first tested a single temperature model (1-$T$ {\tt VAPEC}) but were not able to reproduce the observation, thus the plasma is measurably not isothermal. We then defined a second temperature component (2-$T$ {\tt VAPEC}) with variable Fe abundance for each component and show the result in the top panel of Fig.~\ref{spec}, together with the combined MOS1+2 and PN spectra. The model from each temperature component is shown with dotted lines (for clarity only for MOS1+2). This model already reproduces the data fairly well with a value of $\chi^2_{\rm red}= 1.2$ (at 3076 degrees of freedom). The Fe\,{\sc xxv} lines are 6.7\,keV can only be reproduced if the Fe abundance of the hot component is increased. Meanwhile, the abundance of the cooler component is solar, although with large errors. This means that the hot and cool emission could come from different regions with different metallicity. We added a Gaussian component at 6.4\,keV with zero line width, thus only broadened by the instrument to determine whether there is any significant flux produced by Fe\,{\sc i}. While some flux results, there is not really a line-like features. The inset of Fig.~\ref{spec} shows the PN spectrum around the Fe lines with the 2-$T$ {\tt VAPEC} model (blue line) and the additional emission in the Gaussian (light blue-shadings). The vertical dotted lines mark the energies of the four transitions listed in the left. Two nearby Fe\,{\sc xxv} lines  appear to be resolved with a red-shifted and a blue-shifted component in the PN spectrum, but the model demonstrates that the energy resolution is not sufficient for such a conclusion.

The model parameters are given in the top part of Table~\ref{models}. The values of emission measure, log($VEM$) indicate that the hotter component dominates with a factor 10 over the cooler component. Both temperatures are typical of coronal plasma, and the 2-$T$ {\tt VAPEC} model can thus be considered representative of a coronal plasma with a large range of temperatures.

Closer inspection of the high-energy tail of the 2-$T$ {\tt VAPEC} model in Fig.\,~\ref{spec} (and the inset) suggests that the model systematically underpredicts the continuum in the energy range $\sim 3-7.3$\,keV. This may be an indicator for the high-temperature component to be highly absorbed. This could occur if the high temperature component originates from the boundary layer, thus deeply embedded at the bottom of the accretion disc. To test this scenario, we modify the $2-T$ {\tt VAPEC} model by introducing an additional absorber that acts only on the hot component while the main absorber acts on the entire emission. Since absorption within the accretion disc may not be uniform, we chose a partial absorber. The best-fit parameters are listed in the middle part of Table~\ref{models} while the spectral model is compared with the data in the bottom part of Fig.~\ref{spec}. The lower temperature component is identical in the two models while the high-temperature component yields a lower value in the partial absorber model. The Fe abundance does not need to be modified from solar for this model, indicating that the abundance effect in the 2-$T$ {\tt VAPEC} model may not be real.
The overall value of $N_{\rm H}$ is closer to the interstellar value in the partial absorber model than in the $2-T$ model where it is slightly higher. If ignoring the uncertainties in the interstellar value of $N_{\rm H}$, one might argue that the partial absorber model is more realistic, but the difference is marginal.

Finally, we test the cooling flow model that \cite{mukai2003} have applied to CV spectra after noticing that their grating spectra seem consistent with a multitemperature thermal plasma with a relatively flat emission measure distribution. The flatness of this distribution can indicate an isobaric cooling flow, which assumes that the gas releases all of its energy in the form of optically thin radiation as it cools in a steady-state flow. The optically thin radiation can be modelled with the {\tt APEC} model (see above), and the  {\tt xspec}  model {\tt mkcflow} assumes the flow as an interpolation between a minimum and a maximum {\tt APEC} temperature. The {\tt mkcflow} also has a flavour in which abundances can be modified individually but only for the full flow. The normalization parameter directly gives the total mass flow rate.

The {\tt mkcflow} model has originally been developed for Galaxy Clusters, for which a parameter is needed to red-shift the model. In addition to red-shifting, the parameter $z$ is used to get the flux from the model via distance/$z$. Owing to this dual purpose of $z$, the computation of the mass flow rate from the normalization fails for $z=0$, and for our Galactic source without the need for a red shift, we fix $z$ at a small but non-zero value, $z=5\times 10^{-8}$. This is equivalent to a Doppler shift of 0.015 km\,s$^{-1}$, which is negligible.

The best-fit {\tt mkcflow} model is illustrated in Fig.~\ref{speccflow}, and the parameters are listed in the bottom part of Table~\ref{models}. The inset of Fig.~\ref{speccflow} shows that the Fe lines are underestimated, even when leaving the Fe abundance free to vary. For the $2-T$ {\tt VAPEC} model, a better reproduction of the Fe lines could be achieved because we could modify the abundances of the two components separately. The Fe abundance of the low-temperature component will be driven by the Fe L-shell lines around 1\,keV while the Fe abundance of the high-temperature component will be driven by the K-shell lines at 6.7\,keV. Since the Fe L-shell lines are much stronger, we believe the Fe abundance in the {\tt mkclow} model is driven by the L-shell lines, leading to an overall lower Fe abundance and thus underprediction of the K-shell lines.
The {\tt mkcflow} model also is a poor representation of the energy range $0.7-1$\,keV, however, the overall value of $\chi^2$ is sufficiently small so this model is not necessarily unacceptable compared to the other two models.
\begin{figure}
\resizebox{\hsize}{!}{\includegraphics{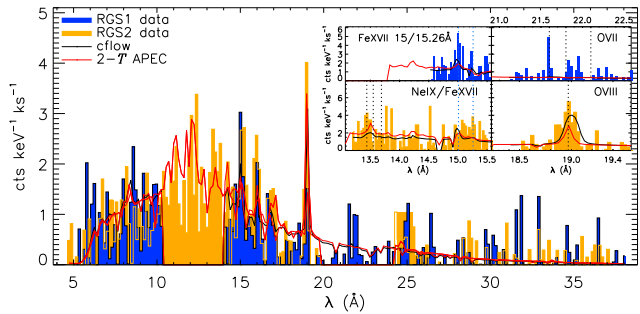}}
\caption{\label{rgsmodel} Comparison of the $2-T$ {\tt VAPEC} and {\tt mkcflow} spectral models (Table~\ref{models}) with the RGS spectra after folding the model through
the RGS1/2 responses. The instet zooms into specific lines as labeled and marked by vertical dotted lines.}
\end{figure}

Although the RGS spectra are not well-enough exposed for quantitative analyses, we can compare the results from the EPIC cameras with the RGS spectrum, and in Fig.~\ref{rgsmodel} we show the RGS1 and RGS2 count spectra in comparison with the $2-T$ {\tt VAPEC} and {\tt mkcflow} spectral models. We have not re-fitted nor re-normalized the models, thus only converted the models to equivalent count spectra by folding them through the spectral responses of the RGS. The main plot shows good agreement of the continuum (a result of the quality of the cross calibration) while the inset shows how much the models agree with selected emission lines. The strongest line, O\,{\sc viii} (18.97\,\AA, bottom right inset) is well reproduced by both models, slightly better with the {\tt mkcflow} model. The O\,{\sc vii} triplet (21.6/21.8/22.1\,\AA, top right inset) is not at all reproduced, suggesting that both models may be missing some cooler plasma. The O\,{\sc vii} forbidden line (22.1\,\AA) seems not detected while the intercombination line (21.8\,\AA) might be present. If this is real, the cooler plasma would have a density above $\sim 10^{10}$\,cm$^{-3}$, higher than typical coronal plasma and could thus come from the boundary layer.

In conclusion, the best fit to the spectra has been obtained with a cool optically thin plasma model plus a partially absorbed hotter component. This model can be interpreted as a coronal component with a temperature $\sim 1$\,keV ($1.2\times 10^7$\,K) and a hotter emission component of $\sim 6$\,keV ($7\times 10^7$\,K) from the boundary layer. In this picture the boundary layer is hidden behind semi-transparent ($\sim 30$\% transparency) plasma with a hydrogen column density of $N_{\rm H}\sim 4\times 10^{22}$\,cm$^{-2}$ while absorption of the coronal plasma is consistent with the interstellar medium. In terms of emission measure, the boundary layer contributes 90\%. However, the other models also lead to acceptable fits, also allowing the possibility that we are either seeing a multi-temperature coronal plasma or a cooling accretion flow, although the cooling flow model requires somewhat higher $N_{\rm H}$ than interstellar.

\begin{table}
\begin{flushleft}
\renewcommand{\arraystretch}{1.1}
\caption{\label{models}Models to EPIC/MOS1+2 spectrum}
{%\scriptsize
\begin{center}
\begin{tabular}{llr}
\hline
Param. & Unit & Value$^a$\\
\hline
\multicolumn{3}{l}{2-$T$ {\tt VAPEC} model}\\
$N_{\rm H}$   & $10^{20}$\,cm$^{-2}$\dotfill & $6.4 - 7.2$ \\
k$T_1$        & keV ($10^6$K)\dotfill       &  $0.98 - 1.02$ (11.4-11.8)\\
Fe abundance$_1$ & solar\dotfill & 1.1-9.0\\
$\log(VEM_1)^c$ & cm$^{-3}$\dotfill & $53.38 - 53.43$\\
k$T_2$        & keV ($10^6$K)\dotfill    &  $7.5 - 8.3$ (87.0-96.3)\\
Fe abundance$_2$ & solar\dotfill & 1.3-1.5\\
$\log(VEM_2)^c$ & cm$^{-3}$\dotfill & $54.49 - 54.50$ \\
Line energy       & keV\dotfill    & 6.4\\
line flux         & $10^{-14}$\,erg\,cm$^{-2}$\,s$^{-1}$\dotfill & $1.9 - 3.2$\\
band flux$^b$ &$10^{-12}$\,erg\,cm$^{-2}$\,s$^{-1}$\dotfill & $2.31 - 2.35$\\
band $L_X^{b,c}$     & $10^{30}$\,erg\,s$^{-1}$\dotfill    & $5.5-5.6$\\
$\chi^2_{\rm red}$ ($dof$) &\dotfill  & 1.17 (3073)\\
\hline
\multicolumn{3}{l}{2-$T$ {\tt VAPEC} model with partial absorber}\\
$N_{\rm H}$   & $10^{20}$\,cm$^{-2}$\dotfill & $5.8 - 6.8$ \\
k$T_1$        & keV ($10^6$K)\dotfill       & $0.98 - 1.02$ (11.4-11.8)\\
$\log(VEM_1)^c$ & cm$^{-3}$\dotfill & $53.37 - 53.42$\\
$N_{\rm H,partial}$   & $10^{22}$\,cm$^{-2}$\dotfill & $2.7 - 5.5$ \\
\hfill fraction & \%\dotfill & $29-36$\\
k$T_2$        & keV ($10^6$K)\dotfill       & $5.7 - 6.4$ (66-74)\\ 
$\log(VEM_2)^c$ & cm$^{-3}$\dotfill & $54.59 - 54.63$\\
$\chi^2_{\rm red}$ ($dof$) &\dotfill  & 1.15 (3074)\\
\hline
\multicolumn{3}{l}{{\tt mkcflow} model}\\
$N_{\rm H}$   & $10^{20}$\,cm$^{-2}$\dotfill & $6.8 - 7.7$ \\
k$T_{\rm low}$   & keV ($10^6$K)\dotfill  &  $0.12 - 0.13$ ($1.4 - 1.5$)\\
k$T_{\rm high}$  & keV ($10^6$K)\dotfill  &  $21.3 - 23.4$ (247.2 - 271.5)\\
$\dot{M}$        & $10^{-12}$\,M$_\odot$\,yr$^{-1}$\dotfill & $6.8 - 7.1$\\
Line energy & keV.....& 6.4\\
Line flux & $10^{-14}$\,erg\,cm$^{-2}$\,s$^{-1}$ & 2.1-3.4\\
band flux$^b$ &$10^{-12}$\,erg\,cm$^{-2}$\,s$^{-1}$\dotfill & $2.2 - 2.3$\\
band $L_X^{b,c}$     & $10^{30}$\,erg\,s$^{-1}$\dotfill    & $5.24-5.48$\\
$L_{\rm bol}^d$  & $10^{30}$\,erg\,s$^{-1}$\dotfill    & $26.5 - 29.2$\\
$\chi^2_{\rm red}$ ($dof$) &\dotfill  & 1.23 (3075)\\
\hline
\end{tabular}
\end{center}
}

$^a$90\% uncertainty ranges\\
$^b0.3-10$\,keV\\
$^c$assuming distance 500\,pc \citep{hoard2004}\\
$^d$from $L=8\times10^{34}\left(\frac{M_{\rm wd}}{{\rm M}_{\odot}}\right) \left(\frac{\dot{M}}{10^{-8} {\rm M}_{\odot} yr^{-1}}\right)\left(\frac{R_{\rm wd}}{5\times 10^8 cm}\right)^{-1}$ with M$_{\rm WD}=0.722$\,$M_{\odot}$ and R$_{\rm wd}=0.01067$\,R$_{\odot}$ \citep{hoard2004}.\\
%$^e$fixed to interstellar value from absorption map
\renewcommand{\arraystretch}{1}
\end{flushleft}
\label{fit_param}
\end{table}

\section{Timing analysis}
\label{timing}

\subsection{Power density spectra}

The goal of the timing analysis is to search for characteristic break frequencies $L_{\rm i}$ that have been detected in the optical PDS determined the from long-term light curve observed with \kepler\ (\citealt{scaringi2012b}).
\begin{table}
\caption{Mean values of the observed break frequencies $L_{\rm i}$ as marked in table~1 of \citet{scaringi2012b} with standard deviation of the sample.}
\begin{center}
\begin{tabular}{lccr}
\hline
\hline
$L_{\rm i}$ & log ($L_{\rm i}$/Hz) & $L_{\rm i}$ & log ($L_{\rm i}$/Hz)\\
\hline
$L_1$ & $-3.01 \pm 0.06$ & $L_3$ & $-3.86 \pm 0.05$\\
$L_2$ & $-3.39 \pm 0.04$ & $L_4$ & $-4.29 \pm 0.11$\\
\hline
\end{tabular}
\end{center}
\label{kepler_pds_param}
\end{table}

\begin{figure}
\includegraphics[width=78mm,angle=-90]{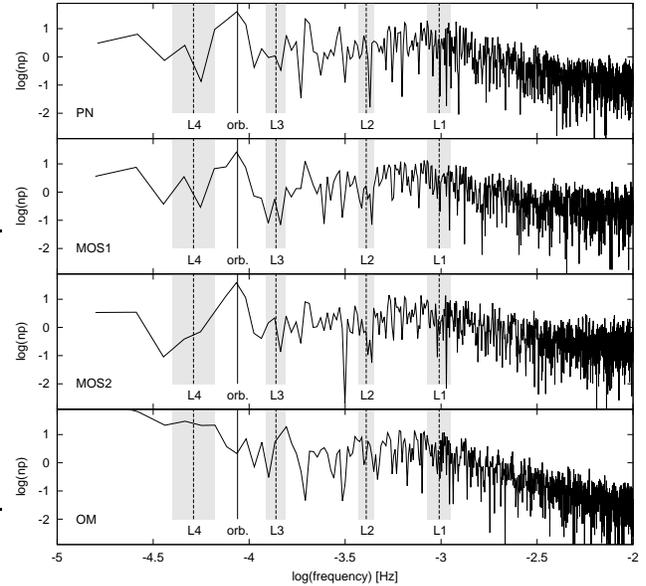}
\caption{PDS of X-ray and UV data taken from different instruments (marked as label in lower left corner). The vertical dashed lines indicate the observed \kepler\ break frequencies and the shaded areas are the error intervals from Table~\ref{kepler_pds_param}. The vertical solid line is the orbital frequency from \citet{skillman1995}.}
\label{pds_all}
\end{figure}

For the timing analysis and PDS calculation we applied the Lomb-Scargle algorithm (\citealt{scargle1982}). This method is particularly adequate for the OM data because it can handle gaps in the light curve. The high and low frequency limits within which the PDS can be studied are determined by the light curve characteristics. The high frequency end is usually limited by the white noise or PDS power rising to the Nyquist frequency. The lower frequency limit is usually determined by the duration of the light curve. The periodograms derived from the individual instruments are depicted in log-log scale in Fig.~\ref{pds_all}. The white noise starts to be noticeable around a frequency of ${\rm log}(f) = -2.4$\,Hz for the X-ray data, but for higher frequencies in the case of UV data. Therefore, we adopted the frequency of ${\rm log}(f) = -2.4$\,Hz as an upper limit for subsequent analysis.

Fig.~\ref{pds_all} reveals an apparent break frequency around $L_1$ in all instruments. Furthermore, the presence of the orbital frequency can clearly be seen in all PDSs except in the UV. While there is no obvious indication for other $L_{\rm i}$ signals in X-rays, the OM data show some increased power around $L_3$ and $L_4$.

For a more detailed study we calculated the PDSs as we have done for the dwarf nova RU\,Peg (\citealt{dobrotka2014}). The light curve is divided into several subsamples, and the corresponding periodograms in log-log scale of every subsample are averaged in order to get the PDS (\citealt{papadakis1993}). The main motivation of this method to average out random features in the individual periodograms, while only real intrinsic PDS features remain. Important is the basic rule, the more subsamples are used, the lower is the scatter in the PDS, while on the other hand, the studied frequency interval becomes narrower with shorter subsamples as the low frequency limit then increases. We subsequently binned the PDS into equally spaced bins in order to reduce the scatter even more and the mean values with the errors were fitted with a model. As bin error estimate we used the standard deviation because it describes the intrinsic PDS scatter. All PDS points within the bin were used for standard deviation calculation. The selection of binning frequency interval and number of divided light curve subsamples we based on $\chi^2_{\rm red}$ and on visual inspection of the binned PDS details. It was clear from the beginning that the binned PDS has a multicomponent shape. Therefore, for fitting we used a multicomponent broken power law consisting of 4 linear functions with three break frequencies. Finally, we divided the OM data into 5 and PN (as the most relevant for timing analysis) into 6 light curve subsamples, with higher binning resolution in the OM data (because of considerably larger count rate), in order to get a solution with $\chi^2_{\rm red}$ closer to 1. For fitting, we used the {\tt Gnuplot}\footnote{http://www.gnuplot.info/} software, yielding best-fit values of break frequencies with standard errors calculated from the variance-covariance matrix.

It is worth noting, that we are comparing quantitatively very different data, i.e. \kepler\ PDSs averaged from 5 one-day light curve segments, vs. 5/6 segments of \xmm\ data with much shorter duration. Therefore, our investigation of the \xmm\ observation is not adequate for detailed studies of low frequencies and thus we concentrate mainly on the break frequencies $L_1$ and $L_2$.

\begin{figure}
\includegraphics[width=54mm,angle=-90]{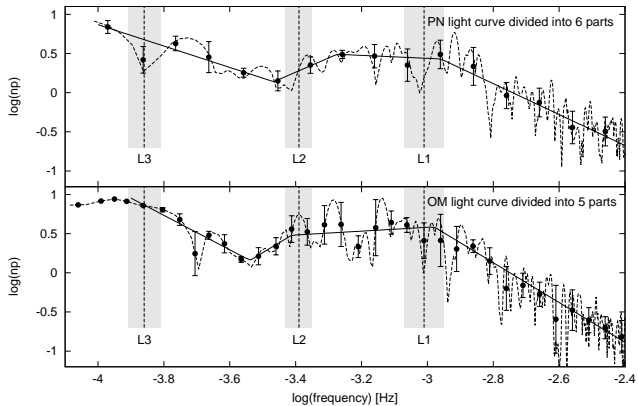}
\caption{PDSs (dashed line) and mean binned PDSs (points with error bars) with broken power law fits (solid line) from PN and OM data. The vertical lines indicate the observed \kepler\ break frequencies and the shaded areas are the error intervals from Table~\ref{kepler_pds_param}.}
\label{pds}
\end{figure}

Resulting best cases with PDS, with binned data and fitted broken power laws are depicted in Fig.~\ref{pds}. The resulting fitted PDS parameters with standard errors are summarized in Table~\ref{xmm_pds_param}.
\begin{table}
\caption{Measured break frequencies from PN ($f_{\rm b1}$) and OM ($f_{\rm b2}$) data with standard errors. The number of light curve subsamples is marked as "nlcs".}
\begin{center}
\begin{tabular}{lcccr}
\hline
\hline
Instrument & log($f_{\rm b1}$/Hz) & log($f_{\rm b2}$/Hz) & $\chi^2_{\rm red}$ & nlcs\\
\hline
PN & $-3.28 \pm 0.07$ & $-2.96 \pm 0.14$ & 0.53 & 6\\
OM & $-3.41 \pm 0.10$ & $-2.98 \pm 0.04$ & 0.73 & 5\\
soft PN & $-3.40 \pm 0.11$ & $-2.99 \pm 0.03$ & 2.18 & 10\\
hard PN & $-3.40 \pm 0.09$ & $-3.04 \pm 0.02$ & 1.29 & 10\\
\hline
\end{tabular}
\end{center}
\label{xmm_pds_param}
\end{table}
The break frequency $L_1$ can clearly be seen in both PN and OM PDSs. All fitted values of the searched break frequency (Table~\ref{xmm_pds_param}) agree well with the observed $L_1$ frequency (Table~\ref{kepler_pds_param}) within the errors. Furthermore, a plateau is seen between $L_1$ and $L_2$ in both PDSs. For OM data the power decrease toward lower frequencies starts at a break consistent with the \kepler\ $L_2$ value. For the PN PDS this break is slightly higher, but all values (PN, OM, \kepler\ $L_2$) agree within the errors, albeit PN and $L_2$ agree only marginally (the error extrema are equal). An additional power increase and possible break close to \kepler\ $L_3$ value is seen in the OM data, although still speculative, but with no indication in PN PDS.

%To investigate the low frequency part of the PDS, another PDS calculation with a smaller number of light curve subsamples is needed. Fig.~\ref{pds_2} shows a 4 subsample case. Such smaller number of subsamples yields a more scattered PDS with larger error bars resulting intuitively in lower $\chi^2_{\rm red}$. Lower frequency bin would reduce the errorbars, but also increase the scatter in the PDS, which is counterproductive. In general, this case appeared to not be adequate for a quantitative study and fitting, and so is the even smaller number of light curve subsamples. From Fig.~\ref{pds_2} we can just conclude that the PDSs show similar characteristics as in 5 subsample case, with a more pronounced $L_3$ break in OM PDS.
%
%\begin{figure}
%\includegraphics[width=54mm,angle=-90]{pds_2.eps}
%\caption{The same as Fig.~\ref{pds}, but with lower light curve division (no PDS is shown for clarity). These data appeared to be not suitable for fitting.}
%\label{pds_2}
%\end{figure}
%

To investigate the low frequency part of the PDS, another PDS calculation with a smaller number of light curve subsamples is needed. However, this yields a much more scattered PDS with larger error bars. A lower frequency bin would reduce the errorbars, but also increases the scatter in the PDS, which is counterproductive. In general, this case appeared to not be adequate for a quantitative study and fitting.

\begin{figure}
\includegraphics[width=54mm,angle=-90]{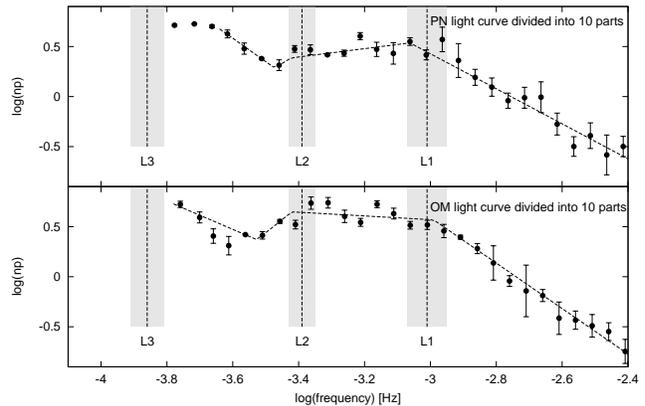}
\caption{Same as Fig.~\ref{pds}, but derived from the division of the light curve into more (10) subsamples. The PDS is not shown for clarity.}
\label{pds_3}
\end{figure}

In the last solution we concentrate on the high frequency part by increasing the number of light curve subsamples. This excludes low frequencies from the PDS, but details in the remaining PDS should be clearer with lower scatter, allowing shorter/finer frequency bins also for the PN PDS. Fig.~\ref{pds_3} shows a 10 subsample case where the presence of a break around $L_2$ is clearer in the PN PDS. The values of $\chi^2_{\rm red}$ are higher (2.0 in the OM PDS) or much higher (10.4 in the PN PDS) than 1 because of smaller errorbars as a natural result of more subsamples used for PDS calculation reducing the scatter.

\begin{figure}
\includegraphics[width=54mm,angle=-90]{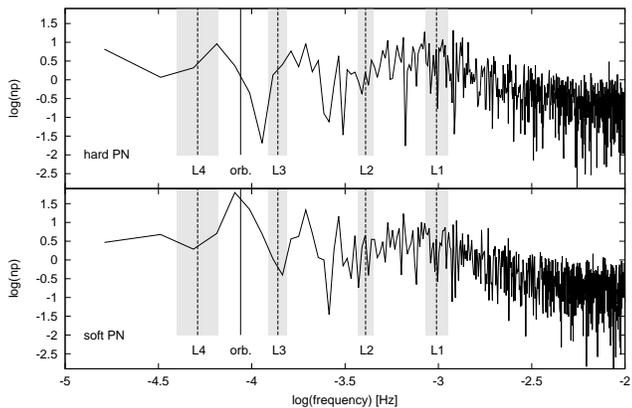}
\caption{The same as Fig.~\ref{pds_all} but for a soft (150-1500\,eV) and a hard (1500-10000\,eV) PN light curve.}
\label{pds_5}
\end{figure}

\begin{figure}
\includegraphics[width=54mm,angle=-90]{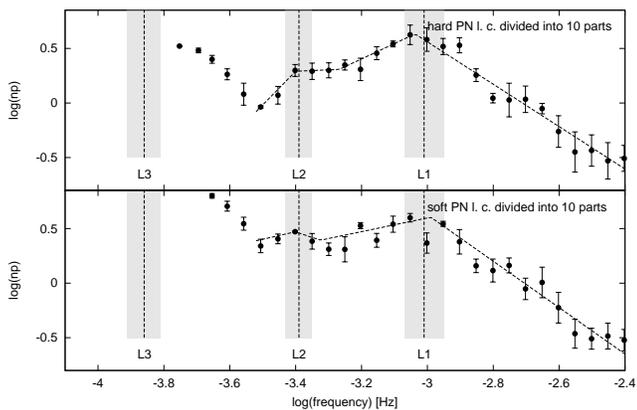}
\caption{The same as Fig.~\ref{pds_3} but for a soft (150-1500\,eV) and a hard (1500-10000\,eV) PN light curve.}
\label{pds_4}
\end{figure}

Based on the results from the spectral analysis (Fig.~\ref{spec}), we created two light curves for a hard and a soft energy band in order to investigate whether the different radiation sources may show different variability patterns. The energy interval for the soft light curve extraction was chosen to roughly agree with the lower temperature plasma component (150 - 1500\,eV) while the hard light curve was extracted from 1500 - 10000\,eV. Non-averaged and non-binned PDSs are depicted in Fig.~\ref{pds_5}. Interestingly, the orbital modulation can only be seen in the soft band while it is absent in the hard X-rays. Averaged and binned PDSs with fits are depicted in Fig.~\ref{pds_4}. For direct comparison we keep the same binning resolution in both cases despite the fact that only one fit yields acceptable $\chi^2_{\rm red}$. Both bands show a clear break frequency coincident with the \kepler\ $L_2$ value, while the break is more pronounced in the hard band. The different PDS behaviour of both bands between $L_1$ and $L_2$ is probably the reason for the slightly higher value obtained from the full PN light curve, i.e. the integrated PDS is slightly deformed by the asymmetry in both bands, while the break agrees well with $L_2$. The fitted PDS parameter values are given in Table~\ref{xmm_pds_param}.

Finally we can conclude that except for the dominant break around the frequency $L_1$, the OM PDS shows an obvious $L_2$ and a possible $L_3$ break, also detected in \kepler\ data, while the PN PDS only shows a break consistent with \kepler\ $L_2$ value.

\subsection{PDS simulations}

While the presence of the $L_1$ break frequency in \xmm\ data is unambiguous, the second break at $L_2$ requires some more justification. For this purpose we simulated artificial light curves following the method by \citet{timmer1995}. The principle is to use two gaussian-distributed random numbers following the input PDS, and use them as the real and imaginary part of the Fourier coefficient. This is done for every frequency and the required time series (synthetic light curve) is obtained by inverse Fourier transform. As input PDS we used PDS parameters from the fits to the real \xmm\ data and the artificial light curves had the same duration and sampling as the original \xmm\ data. We applied the same periodogram calculation, light curve subsample division and binning procedure as for the observations.

\begin{figure}
\resizebox{\hsize}{!}{\includegraphics[angle=-90]{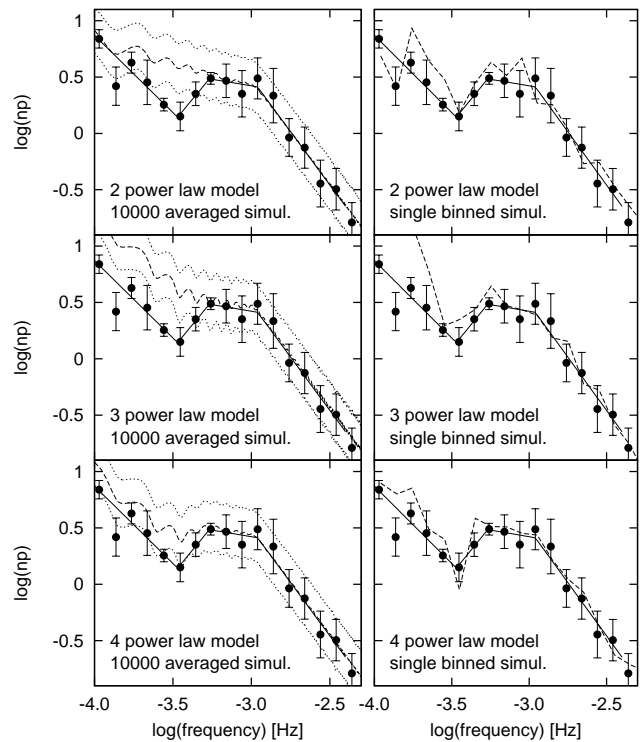}}
\caption{Simulated averaged or binned PDS for various models (see text for details). The simulations are compared to the binned PN PDS (points with errorbars) with the corresponding four-component model (solid line) from the top panel of Fig.~\ref{pds}. The dashed lines are the averaged or binned PDSs and the dotted line is the limit set by the standard deviation of the averaged PDSs.}
\label{pds_simul}
\end{figure}

In Fig.~\ref{pds_simul} we show various simulations compared to the PN binned PDS from the top panel of Fig.~\ref{pds}. We used three PDS models, i.e. a two-component model using two red noises before and after the break frequency at log($f$/Hz) = -2.96, a three-component model, the same as the two component case but the lowest red noise is added, and finally a four-component model as shown by the fit in the top panel of Fig.~\ref{pds}. Therefore, only the latter model comprises the break frequency at log($f$/Hz) = -3.28 ($L_2$ equivalent). The left column of Fig.~\ref{pds_simul} shows the mean PDS calculated from 10000 simulations with standard deviations as assumed error. Apparently the four-component model is the best, although not perfect, i.e. the low-frequency part and the power depression below the $L_2$ break is slightly higher than the original fit used as input for the simulations. This suggests that the random process is important and not every simulated PDS from a light curve divided into 6 subsamples yields the $L_2$ break frequency. In the left column of Fig.~\ref{pds_simul} we display examples of simulated PDSs close to the original fit. Apparently, the power depression below the $L_2$ break frequency can be the result of a random process without the $L_2$ break frequency inherently rooted in the PDS. However, these selected cases in the two and three-component models emerge with lower probability than in the four-component case. Therefore, we can conclude that it is not certain that the break frequency $L_2$ in the observed PN PDS is a real feature, but it is certainly more probable than that it is just a result of a random process.

The presented PDS simulations are ideal for PDS parameters uncertainty derivation directly from the randomness of the red noise. However, the lower break frequency is not present in all simulated PDSs and the fitting procedure did not converge properly in the majority of cases and direct access to the initial parameter estimate was required. Therefore, repeating the simulation process 10000 times, the uncertainty derivation can only be realised for the highest break frequency. We performed simple broken power law fits to the binned PDSs for frequencies higher than log($f$/Hz) = -3.2. A Gaussian fit to the resulting histogram yields a mean value of -3.00 and 1-$\sigma$ parameter of 0.12 as break frequency error. This uncertainty value is very close to the standard error of 0.14 from Table~\ref{xmm_pds_param}. This suggests that the standard errors derived by the {\tt Gnuplot} software are good error estimates.

\subsection{Time delays}

\begin{figure}
\resizebox{\hsize}{!}{\includegraphics[angle=-90]{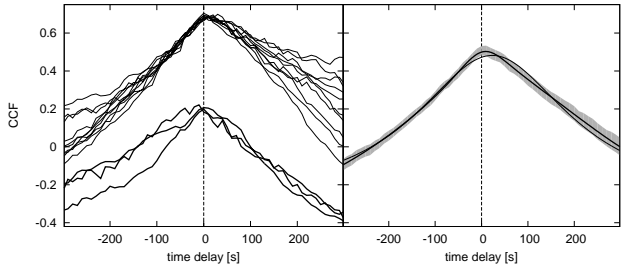}}
\caption{Cross-correlation studies between X-rays and UV. Left panel - 13 CCFs for 13 OM light curves. Different line thicknesses show asymmetry toward opposite time delays. The data were offset vertically for visualization purposes. Right panel - averaged CCF interval using the error of the mean. The thin solid line is a single Lorentzian fit and the thick solid line is the two Lorentzian model. Vertical dashed lines show the zero time lag.}
\label{ccf}
\end{figure}

The UV and X-ray light curves shown in Fig.~\ref{lc} seem to be correlated and in this section we search for time lags between UV and X-rays using a cross-correlation function (CCF) as described in Section~2 of \citet{bruch2015}. Because of the non-continuous nature of the OM light curve, we calculated a separate CCF for each of the 13 continuous OM light curves (Fig.~\ref{ccf} in left panel). Subsequently we calculated a mean CCF which is depicted as a shaded area (the error of the mean is used as an estimate of the error in the CCF) in the right panel of Fig.~\ref{ccf}.

The mean CCF has a maximum around zero time shift but is clearly asymmetric indicating that not all variations are simultaneous or constantly shifted\footnote{A constant time lag would result in a symmetric shape but with a shift of the peak.}. The same indication of the positive time lags (asymmetry or shift of the peak) is seen in almost all individual CCFs except 3 cases, where the indicated time lag is reverse (marked as thick lines in left panel of Fig.~\ref{ccf}). We fitted the mean CCF with a single Lorentzian yielding a poor fit (Fig.~\ref{ccf}, right panel, thin line), while a sum of two Lorentzians yielded an acceptable solution (Fig.~\ref{ccf}, right panel, thick line). The exact values (time lags) of Lorentzian centers depends on the time lag interval over which we performed the fitting. Some time lags are summarized in Table~\ref{ccf_lor_param}\footnote{For time lags larger than 400\,s the CCF is loosing the Lorentzian shape and the wings become constant.}.
\begin{table}
\caption{Central times (time delays) of fitted Lorentzians. $t_{\rm 0}$ is a value from a single Lorentzian model, while $t_{\rm 1}$ and $t_{\rm 2}$ are values for a two-component fit. $\Delta\,t$ is the time lag interval over which the fit has been performed.}
\begin{center}
\begin{tabular}{lccc}
\hline
\hline
$\Delta\,t$ & $t_{\rm 0}$ & $t_{\rm 1}$ & $t_{\rm 2}$\\
(s) & (s) & (s) & (s)\\
\hline
-100 to 100 & $14.6 \pm 1.0$ & $-0.3 \pm 1.9$ & $18.6 \pm 1.3$\\
-200 to 200 & $20.3 \pm 1.2$ & $5.0 \pm 1.2$ & $31.5 \pm 1.7$\\
-300 to 300 & $22.9 \pm 1.2$ & $3.7 \pm 1.0$ & $33.5 \pm 0.9$\\
-400 to 400 & $24.2 \pm 1.2$ & $1.3 \pm 1.5$ & $36.0 \pm 1.2$\\
\hline
\end{tabular}
\end{center}
\label{ccf_lor_param}
\end{table}

\begin{figure}
\includegraphics[width=54mm,angle=-90]{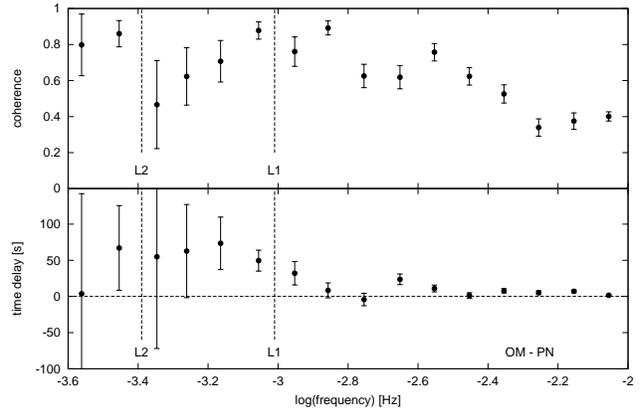}
\caption{Upper panel - coherence between OM and PN light curve as a function of frequency applying the coherence analysis by \citet{nowak1999}. Lower panel - time lag between OM and PN light curve, with PN lagging behind OM for positive values. The horizontal dashed line is the zero time lag. The vertical dashed lines indicate the observed \kepler\ break frequencies from Table~\ref{kepler_pds_param}. The error bars are 3-$\sigma$ uncertainties in both cases.}
\label{time_delay_om-pn}
\end{figure}

Clearly it is not straight forward to derive the time lag from the CCF, but it appears that part of the signal is (almost) simultaneous and part of the X-ray lags behind the UV. Therefore, we performed a more detailed frequency dependent time lag and coherence analysis following the method described by \citet{nowak1999}. The two compared light curves must have the same sampling and number of data points. Therefore, we resampled the OM light curves by linear interpolation in order to synchronise them with the PN data\footnote{The difference is just a few seconds, which is (almost) negligible when comparing both the original and the resampled light curves.} and we filled the gaps with a linear function between two extreme data points of two consecutive OM observations. The coherence and time lags with 3-$\sigma$ uncertainties are shown in Fig.~\ref{time_delay_om-pn}. Clearly, there is a coherence peak around the $L_1$ break frequency, where the positive time lag, in spite of large error bars, confirms the conclusion from the CCF analysis. Also the time delay values are in good agreement with the ones derived from two-Lorentzian CCF fitting ($t_{\rm 2}$ value in Table.~\ref{ccf_lor_param}).

\begin{figure}
\includegraphics[width=66mm,angle=-90]{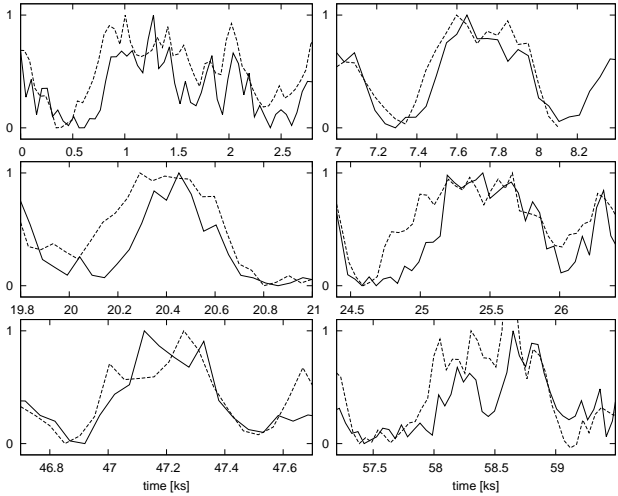}
\caption{Selected light curve segments from Fig.~\ref{lc} showing PN flares (solid thick line) lagging behind the OM flares (dashed thin line). The count rate of both curves is modified in order to get amplitude 1 between the lowest and highest point (except bottom right panel, where we used the second highest point because of a large peak).}
\label{lc_details_1}
\end{figure}

In Fig.~\ref{lc_details_1}, we visualize the behaviour of the PN lag. We selected some examples of UV rising before PN. In order to get comparable light curves, we rescale the selected light curve segments\footnote{Except the flare around 58\,ks (Fig.~\ref{lc} and bottom right panel of Fig.~\ref{lc_details_1}) because there is a very strong peak in PN. Therefore, we choose the second highest peak.}, transforming the count rates to get minima at 0 and maxima at 1. The results suggest that not the whole flare is offset toward earlier time, just the increasing branch shows this characteristic, while the decreasing parts of OM and PN flares are (almost) simultaneous. Therefore, the time delay derived from Figs.~\ref{ccf} and \ref{time_delay_om-pn} is not the true time delay between the peaks, but it represents a time delay of the rising part only.

\begin{figure}
\includegraphics[width=79mm,angle=-90]{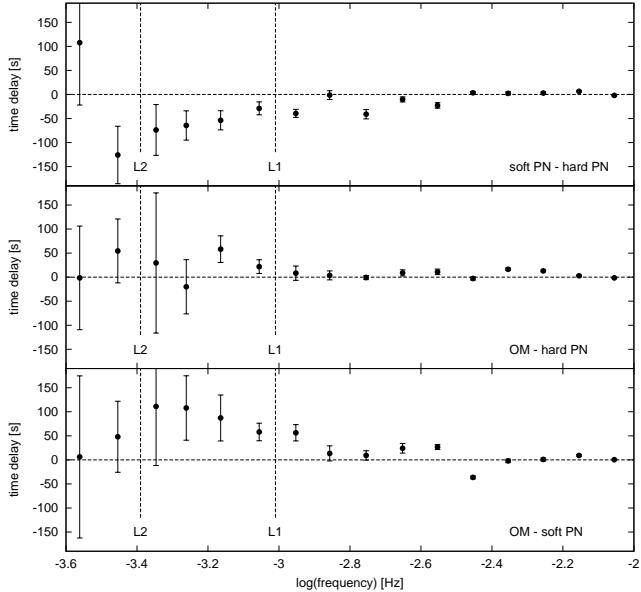}
\caption{The same as lower panel of Fig.~\ref{time_delay_om-pn} but for PN light curve divided into soft and hard band.}
\label{time_delay_others}
\end{figure}

\begin{figure}
\includegraphics[width=36mm,angle=-90]{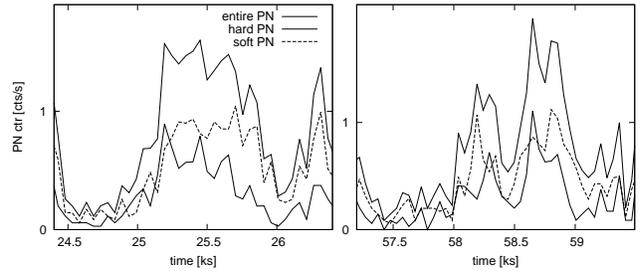}
\caption{The same as Fig.~\ref{lc_details_1} but showing soft PN (dashed line) lagging behind hard PN (solid thin line) at the decline from the flare. The count rate is not modified.}
\label{lc_details_2}
\end{figure}

In order to study the asymmetry in more detail we analysed the time delay between the soft and hard X-ray bands. All corresponding frequency-dependent time lags are visualized in Fig.~\ref{time_delay_others}. It is clear that the soft X-ray emission lags behind the hard emission, while the soft emission lags behind the UV emission for frequencies between $L_1$ and $L_2$, with increasing time delay towards $L_2$. The delay between hard X-rays and UV is not as clear because the data are noisier and many values intersect the zero time lag within the error bars. It is possible that those data are (almost) simultaneous, but a PN lagging behind the OM is still possible mainly around the $L_1$ break frequency.
The situation is depicted in Fig.~\ref{lc_details_2} with two representative examples from Fig.~\ref{lc_details_1}. It is clear how the hard and soft PN count rates rise almost simultaneously, while the hard band is declining earlier. Therefore the mentioned asymmetry/lag between OM and PN is caused mainly by the soft X-ray band. Hard X-rays and UV can thus be more synchronized as suggestive from Fig.~\ref{time_delay_others}.

\begin{figure}
\includegraphics[width=36mm,angle=-90]{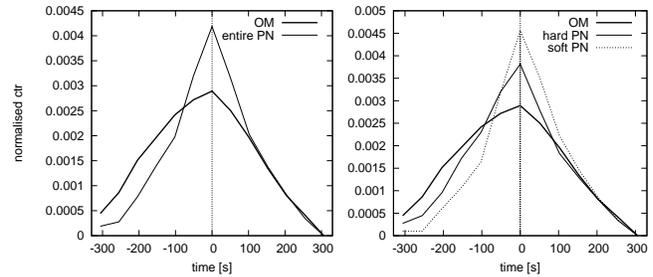}
\caption{Average flare profiles comparison (see text for details). Vertical line is the zero peak time derived from the PN light curve.}
\label{flare_prof}
\end{figure}

For a global study of the mentioned profile features, we analysed the flares in the same way as \citet{negoro2001}, i.e. we averaged several flares in order to get a mean profile. Because of low photon statistics, only the 50\,s binning was adequate for this analysis\footnote{Lower binning does not show smooth flares for the X-ray light curve and a central very thin spike is formed by the averaging instead of a flare.}. The peaks of the flares were defined as the local maxima with 10 data points before and 10 after the maximum in PN light curve, yielding 55 flares. The results are depicted in Fig.~\ref{flare_prof} where the flares were divided by the underlying integrated area in order to make them comparable\footnote{Mainly because of OM flares which have much larger amplitude and count rate than the X-ray}. It is clear from the plots, that the OM is rising before the X-rays, while soft X-rays lag behind the hard band. The simultaneous rise of hard and soft X-rays as suggestive from Fig.~\ref{lc_details_2} is therefore not confirmed. The opposite seems to be present, but all lags are in agreement with the frequency-dependent analysis in Figs.~\ref{time_delay_others} and \ref{time_delay_om-pn}.

Finally, it seems that all peaks are simultaneous. Despite the fact, that lower binning is not adequate for this analysis because of low number of counts, we checked for this simultaneity and it appeared at 30\,s binning. Lower binnings yield a small deviation, but not reliable because of X-ray average flare showing just a very narrow central spike\footnote{It is the result of a flare profile in the original light curve, where the flares are just an accumulation of individual photon spikes.}.

\section{Rms-flux relation}

\begin{figure}
\includegraphics[width=36mm,angle=-90]{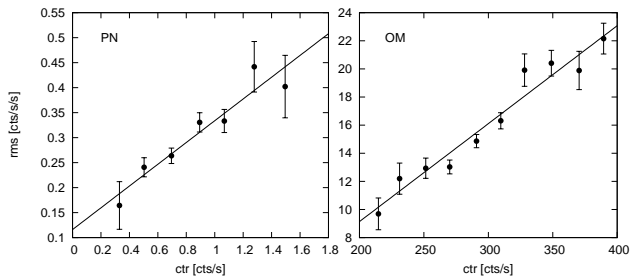}
\caption{Rms-flux relations of PN and OM data. The error bars are the error of the mean and the solid lines are the linear fits to the data.}
\label{rms-flux}
\end{figure}

After the PDS study, we tested whether the \xmm\ data satisfies another typical feature of the flickering in accreting systems, i.e. the rms-flux relation. For this purpose we used the OM light curve samples into 10\,s bins, but for the PN case we needed larger bins of 50\,s because of considerably lower count rate. The resulting rms-flux relation is shown in Fig.~\ref{rms-flux}. The typical linear trend is shown by a linear fit to the data.

\section{Discussion}
\label{discussion}

We present an analysis of \xmm\ observations of the nova-like system MV\,Lyr in order to test the conclusion of \citet{scaringi2014} that the highest break frequency L$_1$ detected in optical \kepler\ data (\citealt{scaringi2012b}) is generated by the central hot optically thin and geometrically thick corona. Such structure is radiating in X-rays, therefore the presence of the break frequency ${\rm log}(L_1) = -3.01 \pm 0.06$\,Hz is expected in X-rays.

\subsection{Timing analysis}

The presence of the searched break frequency around the \kepler\ $L_1$ value can clearly be recognized in all forms of PDS we derived from the X-ray and UV data. The derived value is ${\rm log}(f) = -2.96 \pm 0.14$\,Hz. Furthermore, we found significant indications of the presence of the second optical break frequency $L_2$ found in \kepler\ data. This break frequency with a value of ${\rm log}(f) = -3.41 \pm 0.10$\,Hz is seen in UV data from the OM instrument of \xmm. X-ray data from the PN detector yield only a marginally consistent value of ${\rm log}(f) = -3.28 \pm 0.07$\,Hz, while division of the PN light curve into soft and hard bands yields a value of ${\rm log}(f) = -3.40 \pm 0.09$\,Hz (from the hard band with smaller error and better $\chi^2_{\rm red}$). This break frequency is more pronounced in the hard band.

Finally, we can mention a possible presence of $L_3$ in the OM data while it is not found in the PN data. This would suggest only an optical origin which is in agreement with the disc as a source (\citealt{dobrotka2015a}). But because of low resolution of the lower PDS limit, this conclusion still rests on weak grounds.

\subsection{Time delays}
\label{time_delay}

If the variability characterized by the break frequency $L_1$ is generated by the central corona, the signal must be seen first in X-rays and later reprocessed by the geometrically thin disc (\citealt{scaringi2014}). A time lag, where reprocessed optical/UV radiation lags behind the X-ray is expected. If we assume that such irradiation generates the reprocessed signal with maximal distance from the source to be the outermost boundary of the disc, the light travel time is the expected maximal time lag. The typical radii of discs in CVs are $\sim 10^{10}$\,cm, corresponding to a travel time of 0.3\,s. With the precision of our \xmm\ data we are not able to measure such a small time delay between OM and PN data. Therefore, any discussion about reprocessing time lag is unfortunately not possible.

However, we found some other delays in the \xmm\ light curves. First is the X-ray lagging behind the UV signal between $L_1$ and $L_2$ frequency with a specific characteristics, i.e. X-ray data lagging behind the UV at the rising branch of a flare, while the decrease is simultaneous. Such profile is not new in binary light curves. It was observed by \citet{negoro2001} in the X-ray binary Cyg\,X-1, where the hard band lagged behind the soft on the rising part of a flare, while the decrease was simultaneous.

Furthermore, we showed that the soft X-ray component lags behind the hard component, yielding an even stronger lag of soft X-rays behind the UV signal. The lag of hard X-rays behind the UV is not significant as in other band combinations, but still possible. Finally, we did not find any credible evidence of a lag of entire flares, i.e. the peaks seem to be simultaneous at least any lags are smaller than the bin size.

Because of the uncertain hard X-ray - UV behaviour, we examine three scenarios, i.e. 1) entire X-ray lagging behind the UV, 2) only soft X-ray lagging behind simultaneous UV and hard X-ray (neglecting the small difference), 3) and hard X-ray lagging behind UV, with soft X-ray lagging behind hard X-ray, where all three scenarios have simultaneous peaks.

How to explain the first solution? Let us assume that a turbulent element is formed somewhere in the disc. It needs some time to propagate towards the disc surface and to start to evaporate. Once the evaporation is feeding the corona, the locally enhanced plasma density generates enhanced X-ray emission. Once the turbulent element in the disc dissipates it stops to feed the corona and the response is an immediate decrease of X-ray emission, simultaneously with the turbulent eddy dissipation and local UV radiation. Therefore, the rise delay can be attributed to the turbulent eddy penetration towards the evaporation conditions. Such penetration can also be understood as an expansion which approaches the eddy to the disc surface. The idea of an expanding and cooling sphere, the so-called fireball model, was proposed by \citet{pearson2005} and used by \citet{bruch2015} to explain a typical color delay in CVs flickering. This mechanism does not agree with the interpretation by \citet{scaringi2014} of $L_1$ that the variability is generated by the corona, but it has also problems to explain the detected soft lag behind the hard X-ray.

The second solution is the simultaneous hard X-ray and UV. Two different regions would be responsible for these three radiation bands. A similar behaviour of UV and hard X-rays suggests a region where the hard X-rays are generated by free-free transitions of the hot ionised plasma and the UV is the reprocessing of X-rays by the disc or synchrotron radiation of the same plasma. The soft X-rays with the delay should be generated in a different region of the corona after propagation of the flare generating inhomogeneity/turbulent eddy to different temperature conditions.

Finally, the third scenario can be a combination of the two previous, i.e. an expanding eddy evaporating at the disc-corona transition, with the evaporated matter propagating into a lower temperature region, while the X-rays are reprocessed into UV during the overall process explaining the common decline.

\subsection{Spectral analysis}
\label{spect_anal}

The \xmm\ spectra are well fitted with a 2 temperature {\tt VAPEC} or a {\tt mkcflow} model. Both fits are statistically equivalent suggesting that our interpretation must include a cooler and a hotter source with several points requiring further discussion. For the {\tt VAPEC} model, the hotter component would have a higher Fe abundance which for a partial absorber, applied only to the hot component, would not be necessary.

The first remark is based on a partial absorber that may be associated with the boundary layer between disc and the white dwarf (Fig.~\ref{model}). The soft component may be interpreted as coming from the corona. In such case the detected fast variability can originate in the boundary layer instead of the corona, because the former is the source of 90\% of the emission. At first view, this would disagree with the original assumption based on \citet{scaringi2014} modelling, that the observed \kepler\ $L_1$ break frequency is generated in the corona.

However, \citet{mukai2003} mention the boundary layer as the source of the cooling flow radiation\footnote{Note that the {\tt mkcflow} model is more of an empiric description of boundary layer emission than a rigorous model and might not be suitable for all boundary layers of accretion discs.}. If this localisation is a necessary condition for the cooling flow, a modification is suggested, i.e. X-rays are not generated in the corona nor in the disc-white dwarf boundary layer, but in the boundary layer between the corona and the white dwarf (Fig.~\ref{model}). Such boundary layer is expected, because of different tangential velocities of the corona and the central star. In this case the fast variability can still be generated in the corona by accretion inhomogeneities summed all the way down to the white dwarf surface, where it modulates the X-ray radiation of the boundary layer. Actually this is the principle of the multiplicative accretion process where the inhomogeneities are generated at different radii and transported inwards which modulates the final mass accretion rate (\citealt{lyubarskii1997}, \citealt{kotov2001}, \citealt{arevalo2006}).

The second remark is based on the mass accretion rate derived from the {\tt mkcflow} model, of the order of $10^{-12}$\,M$_{\rm \odot}$\,y$^{-1}$. Such value is acceptable for a dwarf nova during quiescence (\citealt{pandel2005}) but hardly probable for a nova-like in the high state (see Fig.~\ref{lc_aavso}). Furthermore, if the hot component is associated with the boundary layer it should have a temperature of the order of $10^6$ - $10^7$\,K. \citet{godon2011} studied MV\,Lyr in its high state using \fuse\ UV data and derived a boundary layer temperature of $10^5$\,K which is strongly inconsistent with our finding of $10^7$\,K.

However, in a quiescent dwarf nova the inner disc is evaporating and forming a geometrically thick corona ({\citealt{meyer1994}}). All the matter is accreted via this structure, and a UV - X-ray delay is observed. \citet{scaringi2014} proposed a sandwiched model for MV\,Lyr, where a geometrically thick corona is surrounding a geometrically thin disc up to a certain radius, but the geometrically thin disc reaches down to the white dwarf surface. In such a case, only a small fraction of the matter is evaporated and accreted to the central compact object via the corona. This idea is in agreement with our non-detection of a UV - X-ray delay and with the very low mass accretion rate of the order of $10^{-12}$\,M$_{\rm \odot}$\,y$^{-1}$. In such case the disc-white dwarf boundary layer can have the temperature of $10^5$\,K derived by \citet{godon2011}, but the corona-white dwarf boundary layer is much hotter with the temperature found in this paper. The latter is in agreement with the proposed hot boundary layer merged with an accretion-dominated flow (equivalent to corona in our paper) by \citet{balman2014}.

Therefore, both the cool and hot X-ray sources may either be generated in the corona where the hotter component extends closer/down to the white dwarf surface where it would be partially absorbed, or it is associated with the corona-white dwarf boundary layer. In both solutions the mass accretion fluctuations generating the \kepler\ $L_1$ break frequency may be localised in the corona as suggested by \citet{scaringi2014}.

Unfortunately, we cannot distinguish between the case with and without the partial absorber as both models reproduce the \xmm\ data equally well.

In order to compare our observation with the \rosat\ data we used WebPIMMS\footnote{The {\tt WebPIMMS} tool is available at https://heasarc.gsfc.nasa.gov/cgi-bin/Tools/w3pimms/w3pimms.pl} and adopted the (very) simple black-body model of \citet{greiner1998}. We estimated the 0.3-10\,keV band luminosities of all three \rosat\ observations to be $5.87 \times 10^{31}$\,erg\,s$^{-1}$, $4.65 \times 10^{31}$\,erg\,s$^{-1}$, and $1.45 < \times 10^{30}$\,erg\,s$^{-1}$, being the \rosat\ observed luminosity in high state at least a factor $\simeq 8.3$ larger than what obtained by using the \xmm\ data alone and the 2T {\tt VAPEC} model (see Table~\ref{fit_param}).

The results from the spectral analysis of the \swift\ data by \citet{zemko2014} and \citet{balman2014} are inconsistent with the results presented here.
In particular, the models that fit the \xmm\ data are not consistent with the high-state \swift\ spectrum. Since the \swift\ spectra were acquired in a much shorter exposure time than the \xmm\ ones and because of the lower sensitivity, the signal to noise is considerable lower. Especially at high energies, the \swift\ spectra pose little constraint, and we have some doubt that the source characteristics have really changed during the high-state \swift\ spectrum and the new \xmm\ observation. The low signal to noise could introduce random features that are not accounted for in the response
that can lead to inconsistent results when assuming the average response. Note that for \swift, only 'canned' response files are used while for \xmm, a
customized response file is produced based on the most recent calibration, the location of the source on the CCD detector etc. We thus doubt that the source characteristics were really different during the times of observations and give more confidence to the results derived from the EPIC spectra.

\subsection{Comparison with past interpretations}

Part of the X-ray flux is probably generated in a wind from the system (\citealt{balman2014}, \citealt{zemko2014}). This may seem like a contradicting scenario to the one presented in this paper, but in fact it is not. Our corona interpretation is based on a model where inefficient cooling evaporates the corona following \citet{meyer1994}. \citet{balman2014} used archival \rosat\ spectra and fitted them with a combination of cooling flow model with an additional black body component. The latter were not consistent with any optically thick boundary layer (suggested by \citealt{godon2011}). The authors deduced that the boundary layer must be hot with inefficient cooling and an evaporated corona is expected. However, following \citet{meyer1994} the evaporation yields two solutions, i.e. the material is partially accreted on to the white dwarf via a corona, but partially lost via a wind. Therefore, the existence of the corona is directly connected to a wind. Furthermore, X-ray observations of UX\,UMa (\citealt{pratt2004}), an eclipsing VY\,Scl system, show an eclipsed hard (kT $\sim 5$\,keV) component and an uneclipsed soft component. In order to produce eclipses, the hard component must thus be generated close enough to the white dwarf, while the soft component is generated further away from the system to escape eclipses. Moreover, the authors identified the boundary layer as the source of hard X-rays.

MV\,Lyr is a low inclination system without eclipses. If the UV is radiated by the disc, the non detection of the orbital modulation seen in Fig.~\ref{pds_all} is natural. If the hard X-rays are generated in a corona or corona-white dwarf boundary layer, the secondary is not big enough to generate any modulation, but an extended corona in the form of a wind can reach far enough away to be partially occulted. This explains well the presence/absence of orbital modulations of hard/soft X-rays in Fig.~\ref{pds_5}.

\begin{figure}
\resizebox{\hsize}{!}{\includegraphics[angle=-90]{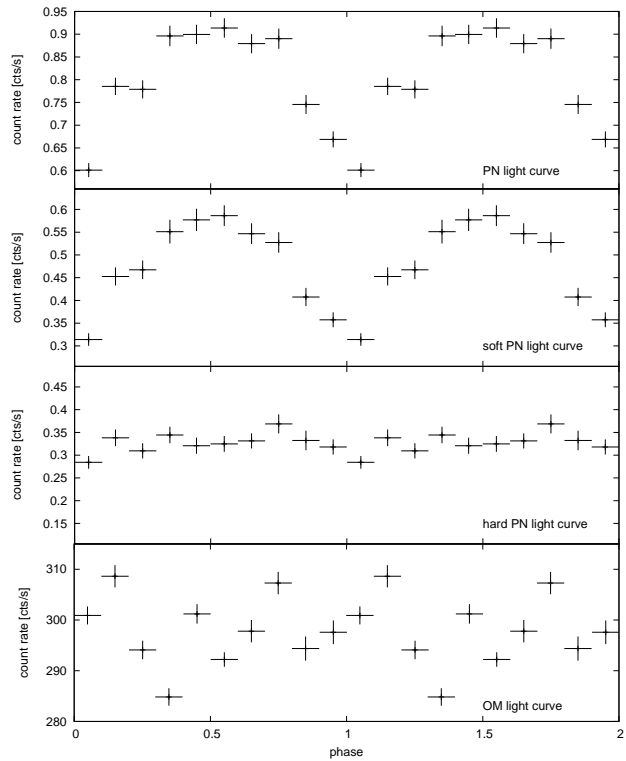}}
\caption{Phased and binned light curves. The y errorbars represent the error of the mean. The x errorbar represents the binned interval. The PN light curves have the same y-axis length for a direct comparison.}
\label{lc_folded}
\end{figure}

\citet{balman2014} folded the X-ray light curve of MV\,Lyr on the orbital period. They did the same for the whole energy interval and using two different intervals for soft (0.3 - 1.0\,keV) and hard (2.5 - 7.0\,keV) light curves. The authors did not detect significant changes in the light curve shape given the statistical errors of the low count rates, and concluded no significant energy dependence in orbital variation. Our modulation finding in Figs.~\ref{pds_all} and \ref{pds_5} suggests the opposite, which is highlighted by our phased and binned soft and hard\footnote{Any possible out of error patterns in hard PN data can be cause by the division of the light curve into two bands which does not exactly separate the spectral components. Therefore, some residual soft X-ray counts are still present in the hard light curve.} \xmm\ data in Fig.~\ref{lc_folded} using an orbital period of 3.19\,h by \citet{skillman1995}.

Finally, \citet{balman2014} pointed out a dilemma in radiation from nova like systems, i.e. the optical and UV accretion rates and luminosities resembling those of dwarf novae in outburst, while their X-ray analysis during the same brightness state with resulting accretion rates and luminosities resemble those in quiescent dwarf novae. The sandwiched model offers a connection between these antagonistic aspects (Fig.~\ref{model}), i.e. the existence of both, the standard disc with an optically thick boundary layer (resembling dwarf novae in outburst, proposed by \citealt{godon2011}), together with a corona with an optically thin hot boundary layer (resembling quiescent dwarf novae, proposed by \citealt{balman2014}).

\subsection{Proposed model}

\begin{figure}
\resizebox{\hsize}{!}{\includegraphics{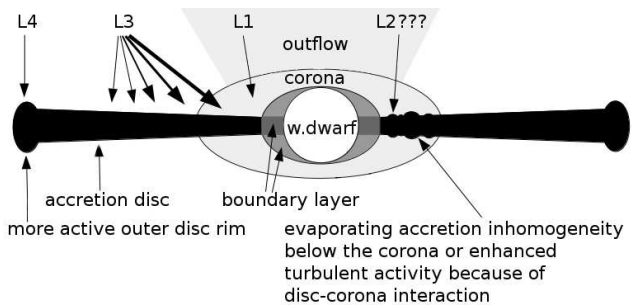}}
\caption{Illustration of the final sandwiched model of MV\,Lyr. The arrows from $L_i$ labels show the origin of the $L_i$ signals. The question mark means that the interpretation of $L_2$ remains uncertain. See text for details.}
\label{model}
\end{figure}

Finally a complex model of the multicomponent PDS shape in MV\,Lyr, that appears the most plausible one to us is illustrated in Fig.~\ref{model}. \citet{scaringi2014} proposed that the highest break frequency $L_1$ detected in Kepler data is generated by a geometrically thick corona, while \citet{dobrotka2015a} found agreement with $L_3$ and the whole geometrically thin disc (the closer to the white dwarf, the more flares are produced) with enhanced activity of the outer disc rim causing the break frequency $L_4$.

However, despite the fact that the model by \citet{dobrotka2015a} is based on a geometrically thin disc, they found the same solution with disc parameters consistent with the geometrically thick disc solution by \citet{scaringi2014} for the frequency $L_1$. \citet{dobrotka2015a} used a statistical method developed by \citet{dobrotka2010} based on various simplifications, while \citet{scaringi2014} used a proper physical method developed by \citet{ingram2013}. Therefore, the geometrically thick nature of the disc solution is more relevant. Besides a coincidence, this can yield an important suggestion, that the flare statistics of a geometrically thin disc is similar to the geometrically thick disc, i.e. the thin disc behaves in a similar way as the corona, and the angular momentum redistribution/gradient, turbulence dimension scales (see \citealt{dobrotka2010} for details) are similar.

The only problematic break frequency is $L_2$. The models of \citet{dobrotka2015a} yield $L_3$ and $L_4$ with $\alpha = 0.1 - 0.4$ and an outer disc radius of 0.5 and 0.9 times the primary Roche lobe. $L_1$ was simulated with a small outer disc radius of $10^{10}$\,cm with $\alpha = 0.9$ as found by \citet{scaringi2014}. But there is a so far ignored solution for $L_2$ in Fig.~4 of \citet{dobrotka2015a} as well, i.e. an outer disc radius equal to the coronal radius and $\alpha = 0.3 - 0.4$. These values of $\alpha$ are consistent with solutions for $L_3$ and $L_4$. In RU\,Peg the low X-ray break frequency was detected also in UV (observed by \xmm) and is interpreted as fluctuations of the inner disc transported to the boundary layer or corona (\citealt{dobrotka2014}), i.e. every fluctuation of mass transfer at the inner disc modulates the UV light and X-ray, because the inner disc is feeding the corona or boundary layer, and every inhomogeneity causes inhomogeneous evaporation or feed of the boundary layer. A similar phenomenon can be applied to MV\,Lyr, i.e. the inhomogeneous mass accretion rate in the geometrically thin disc below the corona modulates the UV and causes inhomogeneous evaporation and subsequent X-ray radiation. The necessary condition is the detection of $L_2$ in UV and X-ray, which seems to be fulfilled, and this speculation would agree with the detected X-ray lag behind the UV (eddy propagation mechanism described in Section~\ref{time_delay}). But the inner disc fluctuations as turbulences were already accounted for in modelling the $L_3$ frequency. Therefore, the inner disc variability should be somehow enhanced to generate an additional component $L_2$. Perhaps the evaporated inhomogeneity is first radiating in X-rays and subsequently they are reprocessed by the geometrically thin disc, which causes the enhanced optical variability of the inner disc below the corona. Perhaps the overall X-ray vs. UV behaviour is a combination of two scenarios; 1) X-ray variability generated by accretion fluctuations in the corona, subsequently reprocessed into optical by the disc, 2) inner thin disc fluctuations radiating in UV and evaporating to the corona where it radiates in X-rays generating additional reprocessed optical signal. Therefore, two processes can be at work leading to the detected complicated band behaviour.

Moreover, a solution where the geometrically thin disc below the corona has enhanced activity because of disc-corona interaction is attractive as well. Different densities and different physical conditions of the disc and corona can yield different tangential velocities of both, and such a velocity gradient is a necessary condition for the Kelvin-Helmholtz instability. Therefore, such transition region can be highly turbulent. A similar interpretation was proposed by \citet{dobrotka2016} in the case of V1504\,Cyg and V344\,Lyr, where a break frequency in the optical PDS is seen during outburst, but absent during quiescence. The disc is developed down to the white dwarf surface in the former case, while it is absent in the latter. The disc-corona interaction during the outburst is proposed to be more turbulent and generating the additional PDS component.

\section{Summary}
\label{summary}

In this paper we analyse new X-ray and UV data of the nova-like system MV\,Lyr taken by \xmm. The results can be summarized as follows.

(i) The X-ray spectra are well fitted with a 2 temperature {\tt VAPEC} or a {\tt mkcflow} model. The {\tt VAPEC} would require different Fe abundance for each temperature component.

(ii) A partial absorber model, applied only to the hotter component, tests the possibility that the hotter component is generated in the boundary layer, while the cooler component originates from the inner geometrically thick corona. The Fe abundance would be the same for both components, giving strong preference to the partial absorber model.

(iii) The derived mass accretion rate and independent boundary layer (between disc and white dwarf) temperature measurements by \citet{godon2011} suggest that the hotter component is probably generated in the inner hot corona or a corona-white dwarf boundary layer. This supports the sandwiched model proposed by \citet{scaringi2014}.

(iv) We confirm the presence of the expected break frequency (${\rm log}(f) = -3.06 \pm 0.02$\,Hz) around the optical signal at ${\rm log}(f) = -3.01 \pm 0.06$\,Hz detected in \kepler\ data (\citealt{scaringi2012b}, \citealt{scaringi2014}). An UV equivalent from \xmm\ data at ${\rm log}(f) = -3.08 \pm 0.03$\,Hz is detected as well.

(v) The second optical break frequency from \kepler\ data of ${\rm log}(f) = -3.39 \pm 0.04$ is present in OM (${\rm log}(f) = -3.34 \pm 0.04$\,Hz) and PN data (${\rm log}(f) = -3.40 \pm 0.09$\,Hz), while it is more pronounced in the harder PN band after light curve decomposition of the X-ray data to soft and hard band based on the two spectral components.

(vi) Soft X-rays are lagging behind both the UV and hard X-rays. The two latter are (almost) simultaneous. Without this decomposition, the entire X-ray flares lag behind the UV, but with probably simultaneous peaks, similar to what is observed by \citet{negoro2001} in Cyg\,X-1. The lags are most pronounced during the rising branch of the flares with an almost simultaneous decline.

(vii) We confirm the sandwiched model suggested by \citet{scaringi2014}, i.e. a central hot geometrically thick optically thin disc is surrounding a geometrically thin and optically thick disc up to a distance of approximately $10^{10}$\,cm from the central white dwarf. The fastest variability around ${\rm log}(f) = -3$\,Hz is generated by accretion fluctuations in the corona and radiated as X-rays. The optical signal is generated by reprocessing of these X-rays by the geometrically thin disc.

(viii) The sandwiched model explains the contradictory resemblance of VY\,Scl systems to dwarf novae, i.e. the corona-white dwarf boundary layer resembles dwarf novae in quiescence, while the underlying geometrically thin disc-white dwarf boundary layer resembles dwarf novae in outburst.

(ix) A complex model is summarized based on additional work of \citet{dobrotka2015a}, where the other three break frequencies detected in \kepler\ data come from the outer disc rim, the whole accretion disc and possibly (just a speculation) from the inner disc region below the corona.

\section*{Acknowledgement}

AD was supported by the Slovak grant VEGA 1/0335/16 and by the ERDF - Research and Development Operational Programme under the project "University Scientific Park Campus MTF STU - CAMBO" ITMS: 26220220179. AAN acknowledges the support by the INFN project TAsP. Furthermore, we acknowledge with thanks the variable star observations from the AAVSO International Database contributed by observers worldwide and used in this research. We thank also Koji Mukai and Polina Zemko for advice and help in spectral modelling, and the anonymous reviewer for valuable comments.

\bibliographystyle{mn2e}
\bibliography{mybib}

\label{lastpage}

\end{document}